\documentclass[lettersize,journal]{IEEEtran}
\usepackage{amsmath,amsfonts}
\usepackage{algorithmic}
\usepackage{algorithm}
\usepackage{amssymb}
\IEEEoverridecommandlockouts
\usepackage{array}
\usepackage[caption=false,font=footnotesize,labelfont=rm,textfont=rm]{subfig}
\usepackage{textcomp}
\usepackage{stfloats}
\usepackage{url}
\usepackage{verbatim}
\usepackage{graphicx}
\usepackage{epsfig} 
\usepackage{hyperref}
\usepackage{cite}
\usepackage{makecell}
\usepackage{xcolor}
\usepackage{fancyhdr}
\fancypagestyle{arxiv}{
	\fancyhf{} 
	\cfoot{
		\parbox{\textwidth}{
			\centering \scriptsize
			\copyright~2026 IEEE. Personal use of this material is permitted. Permission from IEEE must be obtained for all other uses, in any current or future media, including reprinting/republishing this material for advertising or promotional purposes, creating new collective works, for resale or redistribution to servers or lists, or reuse of any copyrighted component of this work in other works. \\
			This work has been published in IEEE Transactions on Vehicular Technology. DOI: 10.1109/TVT.2025.3602807
		}
	}
}

\begin{document}
	
	\title{A Novel 6G Dynamic Channel Map Based on\\ a Hybrid Channel Model}
	
	\author{Tianrun Qi,~\IEEEmembership{Student Member,~IEEE}, Cheng-Xiang Wang,~\IEEEmembership{Fellow,~IEEE}, Chen Huang,~\IEEEmembership{Member,~IEEE}, \\Jiayue Shi,~\IEEEmembership{Student Member,~IEEE}, Junling Li,~\IEEEmembership{Member,~IEEE}, Shuaifei Chen,~\IEEEmembership{Member,~IEEE}, and \\El-Hadi M. Aggoune,~\IEEEmembership{Life Senior Member,~IEEE}.

		\thanks{This work was supported by the National Natural Science Foundation of China (NSFC) under Grants 62394290 and 62394291, the Fundamental Research Funds for the Central Universities under Grant 2242022k60006, the Research Fund of National Mobile Communications Research Laboratory, Southeast University, under Grant 2024A05, the Young Elite Scientists Sponsorship Program by China Association for Science and Technology (CAST) (2022QNRC001), and AI and Sensing Technologies Center, University of Tabuk, KSA, under Grant 1445-200. (\emph{Corresponding Authors}: Cheng-Xiang Wang and Chen Huang.)}
		
		\IEEEcompsocitemizethanks{
			\IEEEcompsocthanksitem T. Qi and J. Shi are with the National Mobile Communications Research Laboratory, School of Information Science and Engineering, Southeast University, Nanjing 210096, China (email: \{tr\_qi, jiayueshi\}@seu.edu.cn).
			\IEEEcompsocthanksitem C.-X. Wang and J. Li are with the National Mobile Communications Research Laboratory, School of Information Science and Engineering, Southeast University, Nanjing 210096, China, and also with the Pervasive Communication Research Center, Purple Mountain Laboratories, Nanjing 211111, China (e-mail: \{chxwang, junlingli\}@seu.edu.cn).
			\IEEEcompsocthanksitem C. Huang and S. Chen are with the Pervasive Communication Research Center, Purple Mountain Laboratories, Nanjing 211111, China, and also with the National Mobile Communications Research Laboratory, School of Information Science and Engineering, Southeast University, Nanjing 210096, China (e-mail: huangchen@pmlabs.com.cn, shuaifeichen@seu.edu.cn). 
			\IEEEcompsocthanksitem H. Aggoune is with Sensor Networks and Cellular Systems Research Center, University of Tabuk, Tabuk 47315, Saudi Arabia (e-mail: hadi.aggoune@gmail.com).
			
	}}


	\maketitle
\thispagestyle{arxiv}
	\begin{abstract}
		In the sixth generation (6G) wireless communication networks, the device density, antenna number, and the complexity of communication scenarios will significantly increase, which brings great challenges for system design and network optimization. By obtaining channel information in advance, channel map has become a promising solution to these challenges in 6G era. However, conventional channel maps cannot be updated in time as physical environment changes. To solve the problem, a novel dynamic channel map (DCM) is proposed in this work. For DCM construction, we further present a ray tracing (RT) and geometric stochastic hybrid channel model (RT-GSHCM), which pre-constructs the DCM offline by RT and updates it online by geometry-based stochastic channel model (GBSM). By this way, the DCM can provide time-varying channel information and channel properties while matintaining accuracy. Next, a channel measurement campaign is conducted, and the measurement results are compared with the RT-GSHCM, RT, and GBSM. The comparison results validate the accuracy of DCM. Meanwhile, the time cost on DCM update is compared with that of conventional channel maps, illustrating the time-efficiency of DCM. Finally, important statistical channel properties of RT-GSHCM are further derived, analyzed, and compared under different configurations of interaction objects in physical environment.
	\end{abstract}
	
	
	\section{Introduction}
	
	The sixth generation (6G) communication network is expected to be a new era of enhanced broadband, massive Internet of things connectivity, and ultra-reliable low-latency communications \cite{background_1}. The goal of 6G is to achieve peak data rates of up to 1 Tbps, significantly improving spectral and energy efficiency compared to the fifth generation communication system, through the utilization of higher frequency bands and artificial intelligence-driven network management for dynamic optimization \cite{background_1,background_2}. Consequently, 6G multiple-input multiple-output (MIMO) communication systems are designed to extend to global coverage scenarios, including various high-mobility communication channels, ensuring effective performance in diverse environments \cite{background_3}. However, the complicated and dynamic communication scenarios, densely distributed communication nodes, and increasing channel dimensions in future 6G MIMO communication systems present numerous challenges. On one hand, existing communication systems widely employ pilot training to acquire channel information, and the pilot overhead increases with the channel dimensions, swiftly consuming the available spectral resources. On the other hand, in variable communication scenarios, the system performance of the existing wireless networks is only realized at 60\%-80\%, leaving significant potential to be tapped \cite{network_1,network_2}.
	
	The studing of the wireless channels is fundamental to the design of communication systems. A profound understanding of channel characteristics largely dictates the quality of system design and network optimization \cite{background_3}, which brings a great demand for accurate channel modeling. Facing the challenges in 6G MIMO communication systems, an effective solution is to explore channel characteristics thoroughly, integrating the challenges into channel studies. By obtaining reliable channel information in advance, it is possible to reduce pilot overhead, optimize wireless network design, manage spectrum resources, improve system performance, and implemente intelligent wireless communication systems.

	As an important concept and tool, channel maps have attracted extensive attention \cite{CKM_1,CKM_2,CKM_3,CKM_4,CKM_5,CKM_6,CKM_7,REM_1,REM_2,REM_3,REM_4,REM_5,REM_6_coushu}. Channel maps provide a detailed description of the wireless communication environment and channel characteristics within a specific geographic area. It provides critical data including interference distribution and parameters of multiple components (MPCs) such as received signal strength, delay, and angular. Existing studies on channel maps include channel knowledge map (CKM) \cite{CKM_1,CKM_2,CKM_3,CKM_4,CKM_5,CKM_6,CKM_7} and radio environment map (REM) \cite{REM_1,REM_2,REM_3,REM_4,REM_5,REM_6_coushu}. CKM is a site-specific database tagged with the locations of the transmitters and receivers, containing channel-related information useful to enhance environment-awareness \cite{CKM_4}. REM is a knowledge database that stores historical data across various domains. It is essential in enhancing the cognitive abilities of radio systems \cite{REM_1}. For both CKM and REM,  there are two solutions for construction, called measurement-based channel map and environment-reconstructed channel map. Measurement-based channel maps store the mapping relationship between measured data and transceiver locations. To construct a comprehensive channel map, machine learning is always ultilized to predict channel information at unmeasured locations \cite{ML_1,ML_2,ML_3,ML_4,ML_5,ML_6}. However, the high cost of channel measurement hinder the map's update as the physical environment changes, thus the channel maps constructed from measurement data cannot adapt to the varied communication scenarios in 6G MIMO communications. Environment-reconstructed channel maps require a two-step construction process. The first step is environmental perception through radar, lidar, camera, or vision-sensing \cite{sense_1,sense_2,sense_3}. Next, channel information are obtained by using the corresponding channel model based on the reconstructed environment. Therefore, an accurate and efficient channel model is critical to construct channel maps precisely.
	
	Unfortunately, current channel models are unable to simultaneously fulfill the demands for accuracy and time-efficiency of environment-reconstructed channel map. Conventional channel models can be divided into deterministic channel model such as ray tracing (RT) and stochastic channel model such as geometry-based stochastic channel model (GBSM) \cite{background_1}. The approach of deterministic channel modeling is to reconstruct the surrounding environment of electromagnetic wave propagation and then physically calculate the attenuation of propagation. As a widely applied deterministic channel model, RT is an electromagnetic field strength prediction algorithm based on geometrical optics and unified diffraction theory \cite{RT_1,RT_2,RT_3,RT_4}. Many existing channel map studies \cite{CKM_2,CKM_3,CKM_4,CKM_5,CKM_6,CKM_7,REM_1,REM_2,REM_3} all used RT to obtain channel information when constructing channel maps. However, RT requires a large amount of geographical information, leading to a high computational complexity. Due to the time-consuming, RT-based channel map cannot support real-time update of channel maps. As for stochastic channel models, GBSM is commonly used for more efficient channel modeling for various communication scenarios \cite{6GPCM}. It has certain geometrical characteristics of multipath, which can describe the impact of environmental scatterers on the channel \cite{GBSM_1,GBSM_2,GBSM_3}. Specially, a 6G pervasive channel model (6GPCM) was proposed to characterize statistical properties of channels at all frequency bands and all scenarios in \cite{6GPCM}. GBSM offers a specific representation of the statistical distribution of obstacles. Compared to RT models, GBSM has low computational complexity. Although considering the impact of scatterers in physical environment, GBSM cannot support a precise channel information output at specific locations due to the randomness.
	
	Meanwhile, hybrid channel models can take advantages of both deterministic channel models and stochastic channel models, and thus, balance the accuracy and time-efficiency \cite{CKM_1}. In \cite{HybridModel_1}, a hybrid channel generation method was developed for unmanned aerial vehicle scenario, where the inter-path parameters were calculated by RT and the intra-path parameters were generated by GBSM. In \cite{HybridModel_2}, a RT-statistical hybrid channel model was developed for the indoor THz channel, which can produce the dominant MPCs by RT and supplement intra-cluster subpaths in the wall-reflection clusters as well as the additional obstacle-reflection clusters. In \cite{HybridModel_3}, a novel hybrid model combining RT and graph theory was introduced for channel modeling in a metropolitan underground environment. The model represented the numerous MPCs resulting from object scattering through a propagation graph approach. This approach involves the creation of scattering points in the vicinity of the interaction points identified by the RT outcomes.
	
	From the literature, conventional channel maps focus on static environment, unable to timely update with the changes of physical environment, limiting their application in 6G MIMO communication systems. In this paper, we present a novel dynamic channel map (DCM), which can provide accurate and time-efficient channel information and channel properties. DCM is first constructed by RT offline, considering the static interaction objects (IOs) in physical environment. Next, based on 6GPCM, DCM can dynamically update channel information online with the changes in physical environment. To construct the DCM, a RT and geometric stochastic hybrid channel model (RT-GSHCM) is proposed. The main contributions of this paper can be summarized as follows:
	\begin{itemize}
		
		\item A novel RT-GSHCM is proposed to construct DCM, where the static and dynamic IOs in physical environment are modeled by RT and 6GPCM, respectively. To describe the weights of static and dynamic components, Rician K-factor $K$ is decomposed into static K-factors $K_S$ and dynamic K-factors $K_D$. By this way, the simulated channels can be accurate and time-varying when physical environment changes.
		
		\item We conduct an urban channel measurement campaign and visualized reconstruct the measurement scenario. By RT offline simulation and 6GPCM online simulation, DCM in the measurement scenario are built. To validate the accuracy and time-efficiency, we first compare the channel properties of DCM, RT, and 6GPCM with those of measured results. Next, time cost on DCM updating is compared with that of conventional channel maps.
		
		\item Statistical properties of the RT-GSHCM are derived and analyzed, including frequency correlation function (FCF), delay power spectrum density (PSD), angular PSD, root mean square (RMS) delay spread, RMS angular spread, RMS Doppler spread, and level crossing rate (LCR). Furthermore, comprehensive comparisons on channel properties under different cluster configurations are conducted.

	\end{itemize}
	
	The rest of this paper is organized as follows. The novel RT-GSHCM is introduced in Section II. Section III gives a brief description of measurement campaign and DCM construction procedure. In Section IV, essential statistical properties of the RT-GSHCM in space, time, and frequency domain are derived.  The evaluation of DCM and the simulation analysis of the RT-GSHCM are given in Section V. Finally, conclusions are drawn in Section VI.
	
	\section{System Model}
\begin{figure}[t]
		\centering
		\includegraphics[width=8cm]{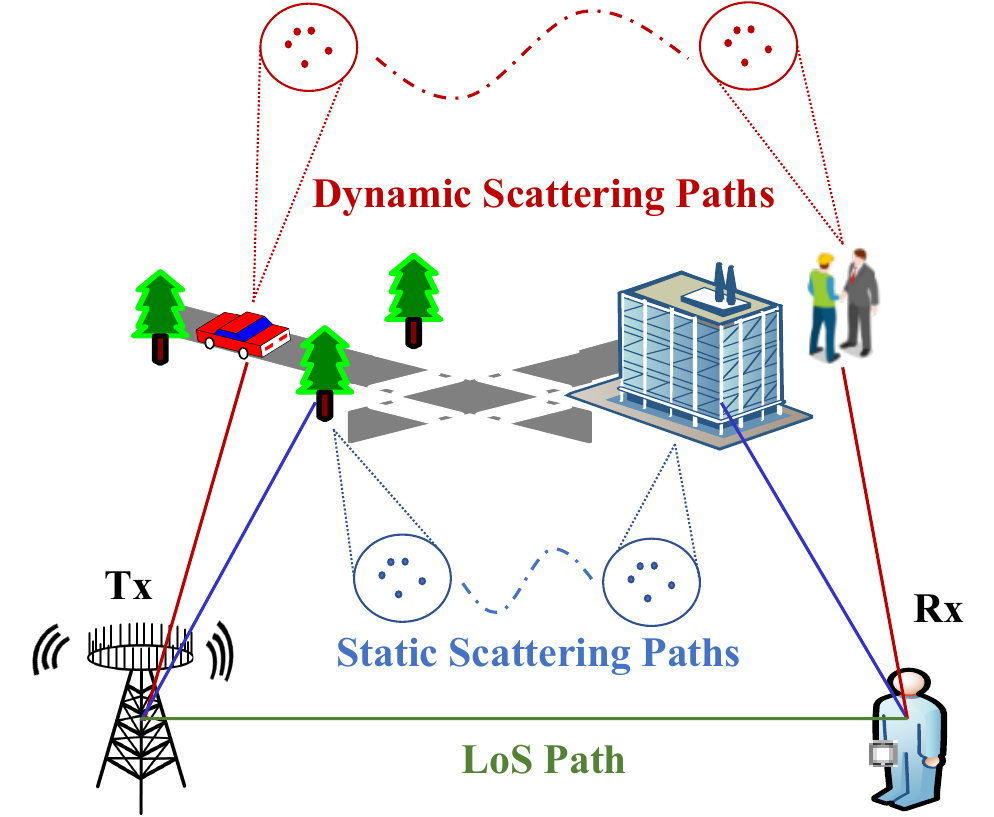}
		\caption{Typical communication environment with static and dynamic IOs in 6G MIMO communications.}
		\label{fig:TypicalCommunicationScenario}
	\end{figure}

  To provide customized high-quality services for different 6G space-air-ground-sea wireless communication scenarios, DCM covers various communication scenarios. Here, we take a typical communication scenario with static and dynamic IOs shown in Fig. \ref{fig:TypicalCommunicationScenario} as an example, to illustrate the procedure of channel generation in RT-GSHCM.
	
	\subsection{General Model Framework}

	\begin{figure*}[b]
	\centering
	\includegraphics[width=17cm]{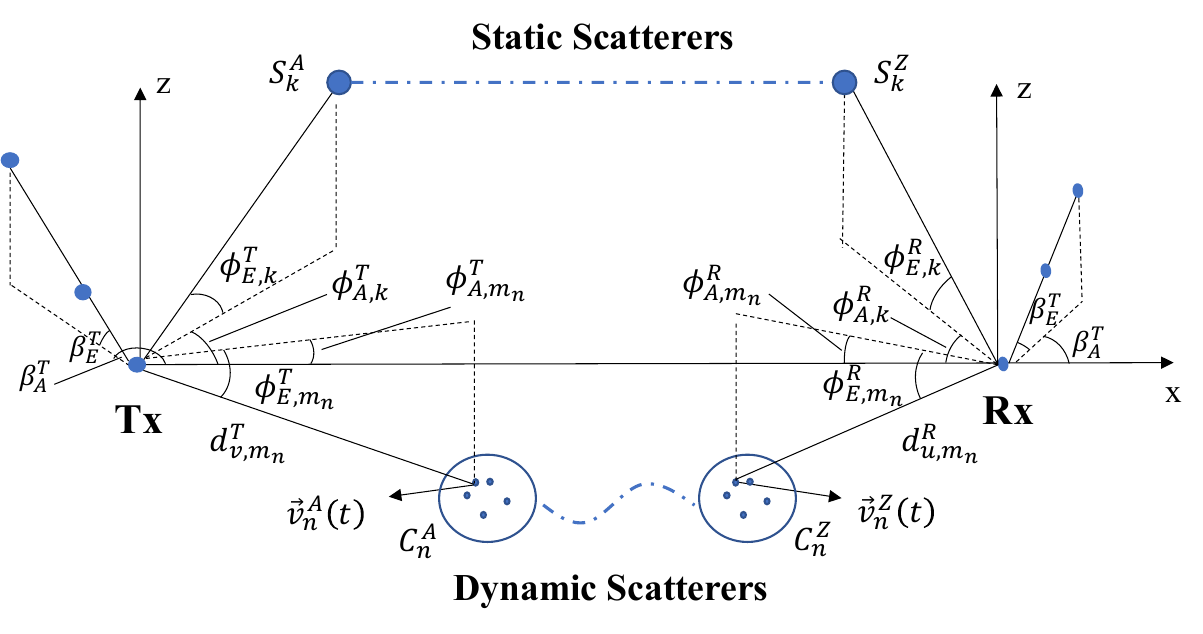}
	\caption{Framework of the proposed RT-GSHCM.}
	\label{fig:Hybrid Channel Model}
\end{figure*}

	The geometrical diagram of the proposed hybrid channel model is presented in Fig. \ref{fig:Hybrid Channel Model}. The numbers of antennas of the transmitter (Tx) and receiver (Rx) sides are denoted as $M_T$ and $M_R$, respectively. $A_v^T$ and $A_u^R$ denote the $v$th $(v = 1,...,M_T)$ and $u$th $(u = 1,...,M_R)$ antenna of Tx and Rx, respectively. Only the $k$th $(k=1,...,K_{vu}(t))$ static scatterer and the $n$th $(n=1,...,N_{vu}(t))$ dynamic cluster are illustrated in Fig. \ref{fig:Hybrid Channel Model}, where $K_{vu}(t)$ and $N_{vu}(t)$ is the number of static and dynamic scatterers between $A^T_v$ and $A^R_u$ at moment $t$. Here, $K_{vu}(t)$ is time-varying because the motion of Tx and Rx. Considering the multi-bounce propagation, the $n$th dynamic cluster is denoted as cluster pairs, including the first-bounce cluster $C^A_n$ at Tx side and the last-bounce cluster $C^Z_n$ at Rx side. The $m$th $(m=1,...,M_n)$ ray in $C_n^A$ is also illustrated in Fig. \ref{fig:Hybrid Channel Model}, where $M_n$ represents the total number of rays in $C_n^A$. For the static IOs in physical environment, line of sight (LoS) and specular contributions are considered as two major wave propagation mechanisms, and the MPCs based on the LoS and specular contributions are defined as static scattering MPC. The propagation through the $k$th static scattering MPC is modeled by RT and denoted as scatterer pairs. The proposed hybrid channel model parameters are summarized in Table~\ref{tab:model parameter}.

	The total channel matrix of the proposed hybrid model is the multiplication of the large scale fading (LSF) and small scale fading (SSF), which can be calculated as
	\begin{equation}
		\begin{split}
			\mathbf{H} = LSF \cdot SSF = \sqrt{PL \cdot SF} \cdot \mathbf{H_{SSF}},
		\end{split}
	\end{equation}
	where $PL$ denotes the path loss caused by the propagation distance between TX and Rx, $SH$ denotes shadowing fading, and $\mathbf{H_{SSF}}$ denotes the SSF matrix, which can be written as 
	\begin{equation}
		\begin{split}
			\mathbf{H_{SSF}} = [h_{vu}(t,\tau,\ell)]_{M_R\times M_T},
		\end{split}
	\end{equation}
	where $\ell$ stores the locations of Tx and Rx in global coordinate and can be expressed as $\ell = \{\ell^T,\ell^R\}$. Contrasting the conventional GBSM relying on the relative distances between Tx and Rx, the RT-GSHCM leverages the transceiver locations within a global coordinate. Consequently, the DCM constructed by RT-GSHCM can provide accuracy channel information according to transceiver locations. $h_{vu}(t,\tau,\ell)$ denotes the channel impulse response (CIR) between $A^T_v$ and $A^R_u$, calculated as the sum of static component and dynamic component
	\begin{equation}
		\begin{aligned}
			h_{vu}(t,\tau,\ell) &= \sqrt{\frac{K_S^{-1}+1}{K_S^{-1}+K_D^{-1}+1}}h^{RT}_{vu}(\tau,\ell)\\ &+\sqrt{\frac{K_D^{-1}}{K_S^{-1}+K_D^{-1}+1}}h^{D}_{vu}(t,\tau,\ell), \label{CIR}
		\end{aligned}
	\end{equation}
	where $h^{RT}_{vu}$ and $h^D_{vu}$ denote the static scattering components and dynamic scattering components, respectively. $K_S$ and $K_D$ denote the ratio of the power of LoS components to the power of static/dynamic scattering components. It is well known that the Rician K-factor $K$ is the ratio of the power of LoS components to the power of non-line of sight (NLoS) components. Therefore, $K$ can be calculated as
		\begin{table*}[t]
		\caption{Definitions of parameters in the RT-GSHCM.}
		\renewcommand{\arraystretch}{1.5}
		\begin{center}
			\begin{tabular}{|c|c|}
				\hline
				\textbf{Parameters} &  \textbf{Definitions} \\ \hline
				$S^A_k,S^Z_k$ & The first and last bounce of the $k$th static scatterer at Tx and Rx\\ \hline
				$S^A_{m_n},S^Z_{m_n}$ & The $m$th scatterer in $C^A_{n}$ / $C^Z_{n}$\\ \hline
				$\beta^T_{E/A},\beta^R_{E/A}$ & The elevation/azimuth angle of the Tx and Rx antenna array\\ \hline
				$\alpha^{A_n}_{E/A}(t),\alpha^{Z_n}_{E/A}(t)$ & The elevation/azimuth angle of moving direction of $C^A_n$ and $C^Z_n$ at time instant $t$\\ \hline
				$\phi^{T}_{E/A,m_n}(\ell),\phi^{R}_{E/A,m_n}(\ell)$ & The elevation/azimuth angle of departure and arrival of $S^A_{m_n}$ and $S^Z_{m_n}$ at initial moment in location $\ell$\\ \hline
				$\phi^{T}_{E/A,k}(\ell),\phi^{R}_{E/A,k}(\ell)$ & The elevation/azimuth angle of departure and arrival of $S^A_{k}$ and $S^Z_{k}$ in location $\ell$\\ \hline
				$v^A_n(t),v^Z_n(t)$ & The absolute value of the velocity of $C^A_n$ and $C^Z_n$ at time instant $t$\\ \hline
				$d^{T}_{m_n}(\ell),d^{R}_{m_n}(\ell)$ & Distance between $A^T_1$ and $S^A_{m_n}$ / $A^R_1$ and $S^Z_{m_n}$ at initial moment in location $\ell$\\ \hline
			\end{tabular}
			\label{tab:model parameter}
		\end{center}
	\end{table*}

	\begin{equation}
		\frac{1}{K} = \frac{1}{K_S}+\frac{1}{K_D}. \label{K}
	\end{equation}
		\begin{figure*}[b]
		\hrulefill
		\begin{align}
			h_{vu}^{L}(\tau,\ell)=&\left[\begin{array}{c}F_{u,V}\left(\phi_{E,L}^{R}(\ell),\phi_{A,L}^{R}(\ell)\right)\\F_{u,H}\left(\phi_{E,L}^{R}(\ell),\phi_{A,L}^{R}(\ell)\right)\end{array}\right]^{\mathrm{T}}\begin{bmatrix} e^{j\theta_{L}^{VV}}&0 \\0&e^{j\theta_{L}^{HH}}\end{bmatrix}\left.\left[\begin{array}{c}F_{v,V}\left(\phi_{E,L}^T(\ell),\phi_{A,L}^T(\ell)\right)\\F_{v,H}\left(\phi_{E,L}^T(\ell),\phi_{A,L}^T(\ell)\right)\end{array}\right.\right] \notag \\
			&e^{j2\pi f_{c}\tau_{vu}^L(\ell)}\cdot\delta\left(\tau-\tau_{vu}^L(\ell)\right) \tag{6} \label{h^LoS}\\
			h_{vu}^{S}(\tau,\ell)=&\sum_{k=1}^{K_{vu}(\ell)}\left[\begin{array}{c}F_{u,V}\left(\phi_{E,k}^{R}(\ell),\phi_{A,k}^{R}(\ell)\right)\\F_{u,H}\left(\phi_{E,k}^{R}(\ell),\phi_{A,k}^{R}(\ell)\right)\end{array}\right]^{\mathrm{T}}\begin{bmatrix}e^{j\theta_{k}^{VV}}&\sqrt{\mu\kappa_{k}^{-1}(\ell)}e^{j\theta_{k}^{VH}}\\\sqrt{\kappa_{k}^{-1}(\ell)}e^{j\theta_{k}^{HV}}&\sqrt{\mu}e^{j\theta_{k}^{HH}}\end{bmatrix} \notag \\
			&\left.\left[\begin{array}{c}F_{v,V}\left(\phi_{E,k}^T(\ell),\phi_{A,k}^T(\ell)\right)\\F_{v,H}\left(\phi_{E,k}^T(\ell),\phi_{A,k}^T(\ell)\right)\end{array}\right.\right]\sqrt{P_{vu,k}(\ell)} \cdot e^{j2\pi f_{c}\tau_{vu,k}(\ell)}\cdot\delta\left(\tau-\tau_{vu,k}(\ell)\right) \tag{7}\label{h^S}
		\end{align}
	\end{figure*}

	We will describe in detail the static channel modeling and how to use the static channel information from the RT method to offline model the accurate static scatterering channel at initial moment. The CIRs of static scattering channel and dynamic scattering channel will be introduced in the next two subsections, respectively.
	
	\subsection{Static CIR}
	For the static environment, RT is used to generate the LoS path and the MPCs of static scatterers, such as buildings, trees, and ground. RT can analyze the electromagnetic propagation process accurately and produce the MPCs' parameters. The static CIR in \eqref{CIR} can be calculated as
	\begin{equation}
		\begin{split}
			h^{RT}_{vu}(\tau,\ell)=\sqrt{\frac{1}{K_S^{-1}+1}}h^{L}_{vu}(\tau,\ell)+\sqrt{\frac{K_S^{-1}}{K_S^{-1}+1}}h^{S}_{vu}(\tau,\ell),
		\end{split}
	\end{equation}
	where $h^{L}_{vu}(\tau,\ell)$ and $h^{S}_{vu}(\tau,\ell)$ denote the LoS component and NLoS component of static CIR, respectively. The calculations of $h^{L}_{vu}(\tau,\ell)$ and $h^{S}_{vu}(\tau,\ell)$ can be expressed as \eqref{h^LoS} and \eqref{h^S} shown at the bottom of this page.
	
	Here, $\{\cdot\}^{\mathrm{T}}$ denotes the transposition operation, $F_{u/v,V(H)}$ stand for the vertical (horizontal) polarization of $A^T_v$ / $A^R_u$. $\kappa_k(t)$ is the cross polarization power ratio, $\mu$ is co-polar imbalance \cite{II-1}, $\theta^{VV}_L$, $\theta^{VH}_L$, $\theta^{VV}_k$, $\theta^{VH}_k$, $\theta^{HV}_k$, and $\theta^{HH}_k$ are the initial phases uniformly distributed between $[0, 2\pi]$. $f_c$ denotes carrier frequency. 

	In \eqref{h^S}, $P_{vu,k}(\ell)$ and $\tau_{vu,k}(\ell)$ are the power and propagation delay of the $k$th ray between antenna $A^T_v$ and $A^R_u$ at location $\ell$. $\phi_{E/A,k}^{T/R}(\ell)$ denotes the elevation / azimuth angle of departure / arrival (EAoD / AAoD / AAoA / AAoD) of the static scattering MPCs between $S^A_{k}$ and $S^Z_{k}$ at location $\ell$. When it comes to the LoS components in \eqref{h^LoS}, $\tau^L_{vu}(\ell)$ is the propagation delay of the LoS path. $\phi_{E(A),L}^{T/R}(\ell)$ is the elevation (azimuth) angle of departure/arrival of the LoS path at location $\ell$. Here, MPC parameters such as $P_{vu,k}(\ell)$, $\tau_{vu,k}(\ell)$, $\tau^L_{vu}(\ell)$, $\phi_{E/A,k}^{T/R}(\ell)$, and $\phi_{E(A),L}^{T/R}(\ell)$ are all provided by RT simulator.

	\subsection{Dynamic CIR}
	 To efficiently characterize the dynamic IOs in physical environment, 6GPCM is used in the dynamic component of the RT-GSHCM \cite{6GPCM}. It can provide pervasiveness, applicability and time-efficiency for the DCM. Note that only NLoS path is considered in dynamic scattering channel. Consequently, the expression is illustrated at the bottom of this page\cite{6GPCM}.
	
	\begin{figure*}[b]
	\hrulefill
	\begin{equation}
		\begin{aligned}
			h_{vu}^{D}(t,\tau,\ell)=&\sum_{n=1}^{N_{vu}(t,\ell)}\sum_{m=1}^{M_{n}}\left[\begin{array}{c}F_{u,V}\left(\phi_{E,m_{n}}^{R}(t,\ell^R),\phi_{A,m_{n}}^{R}(t,\ell^R)\right)\\F_{u,H}\left(\phi_{E,m_{n}}^{R}(t,\ell^R),\phi_{A,m_{n}}^{R}(t,\ell^R)\right)\end{array}\right]^{\mathrm{T}}\begin{bmatrix}e^{j\theta_{mn}^{VV}} & \sqrt{\mu\kappa_{m_{n}}^{-1}(t,\ell)}e^{j\theta_{mn}^{VH}}\\\sqrt{\kappa_{m_{n}}^{-1}(t,\ell)}e^{j\theta_{mn}^{HV}} & \sqrt{\mu}e^{j\theta_{mn}^{HH}}\end{bmatrix} \\
			&\left.\left[\begin{array}{c}F_{v,V}\left(\phi_{E,m_n}^T(t,\ell^T),\phi_{A,m_n}^T(t,\ell^T)\right)\\F_{v,H}\left(\phi_{E,m_n}^T(t,\ell^T),\phi_{A,m_n}^T(t,\ell^T)\right)\end{array}\right.\right]\sqrt{P_{vu,m_n}(t,\ell)} \cdot e^{j2\pi f_{c}\tau_{vu,m_{n}}(t,\ell)}\cdot\delta\left(\tau-\tau_{vu,m_{n}}(t,\ell)\right).  \refstepcounter{equation}\refstepcounter{equation}
		\end{aligned}
	\end{equation}
\end{figure*}
	
	Here, $\kappa_k(t,\ell)$ is the cross-polarization power ratio. $P_{vu,{m_n}}(t,\ell)$ and $\tau_{vu,{m_n}}(t,\ell)$ are the power and propagation delay of the $k$th ray between antenna $A^T_v$ and $A^R_u$ at time instant $t$, where $\tau_{vu,{m_n}}(t,\ell)$ can be calculated as $\tau_{vu,m_n}(t,\ell)=\frac{d^T_{v,m_n}(t,\ell)+d^R_{u,m_n}(t,\ell)}{c}+\tilde{\tau}_{m_n}$, where $c$ denotes the speed of light in free space. $\tilde{\tau}_{m_n}$ denotes the delay of virtual link between $C^{A}_{m_n}$ and $C^{Z}_{m_n}$ as $\tilde{\tau}_{m_n}=\tilde{d}_{m_n}/\tau_{C,link}$, where $\tau_{C,link}$ is a non-negative variable randomly generated according to the exponential distribution. $d^T_{v,m_n}(t,\ell)$ is the distance between $A^T_v$ and $S^{A}_{m_n}$, modeled as $d^T_{v,m_n}(t,\ell)=||\vec{d}^{T}_{m_n}(\ell)+\int_0^t{\vec{v}^{A}_n(t)}dt-\vec{l}^T_v||$. $\vec{d}^{T}_{m_n}(\ell)$ denotes the vector pointing from $A^T_1$ to $S^{A}_{m_n}$ at the initial moment. $\vec{v}^{A}_n(t)$ is the vector of velocity of $C^A_n$. $\vec{l}^T_v$ denotes the vector pointing from $A^T_1$ to $A^T_v$. They can be calculated as

		\begin{align}
			\vec{d}^{T}_{m_n}(\ell)=&d^{T}_{m_n}(\ell) \cdot [ \cos{\phi^{T}_{E,m_n}(\ell)} \cos{\phi^{T}_{A,m_n}(\ell)},\\
			&\cos{\phi^{T}_{E,m_n}(\ell)} \sin{\phi^{T}_{A,m_n}(\ell)},\sin{\phi^{T}_{E,m_n}(\ell)} ]\notag \\
			\vec{v}^{A}_n(t)=&v^{A}_n(t) \cdot [\cos{\alpha^{A_n}_E(t)} \cos{\alpha^{A_n}_A(t)}, \\
			&\cos{\alpha^{A_n}_E(t)} \sin{\alpha^{A_n}_A(t)},\sin{\alpha^{A_n}_E(t)}]\notag \\
			\vec{l}^T_v=&\delta_v \cdot [\cos{\beta^T_E} \cos{\beta^T_A},\cos{\beta^T_E} \sin{\beta^T_A},\sin{\beta^T_E}] \label{l^T}.
		\end{align}

	The variable $\delta_v$ in \eqref{l^T} represents the distance from $A^T_v$ to $A^T_1$, calculated as $\delta_v=(v-1)\delta^T$. Here, $\delta^T$ denotes the antenna interval of Tx.
	
		To enhance the accuracy and practicality of the RT-GSHCM, channel parameters in RT-GSHCM are optimized based on dynamic components prior to deployment. Consequently, the dynamic CIR of RT-GSHCM has parameter values that differ significantly from the purely stochastic 6GPCM. This divergence primarily arises from the distinct parameter optimization processes: RT-GSHCM optimizes parameters specifically for the dynamic IOs identified in the environment, while the 6GPCM parameters are globally optimized to reflect the whole scenario characteristics, including static and dynamic IOs. Hence, RT-GSHCM can better capture location-dependent channel dynamics introduced by specific moving scatterers.
	
		By such 6GPCM-based dynamic components simulation, the proposed RT-GSHCM can update channel parameters such as cluster positions and velocities in real-time with sensors. For scenarios where vehicles or UAVs act as moving scatterers. This update mechanism enables accurate and low-latency channel mapping in highly dynamic environments.
	
	\subsection{Channel Transfer Function (CTF)}
	Making the Fourier transform of $h_{vu}(t,\tau,\ell)$, we can obtain the CTF of the RT-GSHCM as
	\begin{equation}
		\begin{aligned}
			H_{vu}(t,f,\ell) &= \sqrt{\frac{K_D^{-1}}{K_S^{-1}+K_D^{-1}+1}}H^D_{vu}(t,f,\ell)\\
			&+\sqrt{\frac{1}{K_S^{-1}+K_D^{-1}+1}}H^{L}_{vu}(f,\ell) \\
			&+\sqrt{\frac{K_S^{-1}}{K_S^{-1}+K_D^{-1}+1}}H^{S}_{vu}(f,\ell), \label{CTF}
		\end{aligned}
	\end{equation}
	where $H^D_{vu}(t,f,\ell)$, $H^{L}_{vu}(f,\ell)$, and $H^{S}_{vu}(f,\ell)$ are the CTFs of dynamic scattering channel, LoS component, and NLoS component in static scattering channel, respectively. For the convenience of calculation, the antenna polarization is not considered in CTFs. Therefore, the CTFs can be written as
	\begin{align}
		H^D_{vu}(t,f,\ell) =& \sum_{n=1}^{N_{vu}(t,\ell)}\sum_{m=1}^{M_{n}}\sqrt{P_{vu,m_n}(t,\ell)}  \\
		&\cdot e^{-j2\pi \tau_{vu,m_{n}}(t,\ell)(f-f_{c})}\notag\\
		H^{L}_{vu}(f,\ell) =& e^{-j2\pi \tau^L_{vu}(\ell)(f-f_{c})}\\
		H^{S}_{vu}(f,\ell) = &\sum_{k=1}^{K_{vu}(\ell)}\sqrt{P_{vu,k}(\ell)} \cdot e^{-j2\pi \tau_{vu,k}(\ell)(f-f_{c})}.
	\end{align}

	In addition, it is necessary to clarify that the static components generated by RT exhibit fixed amplitudes and phases at a given transmitter-receiver (Tx–Rx) location though consisting of multiple discrete paths. As these components are deterministic and time-invariant within the short timescale relevant for small-scale fading, their complex vector sum can be treated as a single equivalent deterministic component 
\begin{equation}
	H_{static} = \sum_{k=1}^{K_S+1}h^{RT}_{vu} = \sum_{k=1}^{K_S+1} a_k e^{j\phi_k} \triangleq A,
\end{equation}
where $a_k$ and $\phi_k$ denote the amplitude and phase of the $k$-th static MPC, respectively. 

The dynamic multipath component $h_D(t)$, generated by the geometric-based stochastic channel model (GBSM), consists of numerous paths with random amplitudes and phases, which collectively approximate a zero-mean complex Gaussian random process due to the central limit theorem. Here, K-factor has already been considered in $A$ and $h_D(t)$. 

Consequently, the total small-scale fading channel at time $t$ is expressed as:
\begin{equation}
	H(t) = H_{static} + h_D(t) = A + h_D(t).
\end{equation}

Because the deterministic static component $A$ is combined with the zero-mean Gaussian dynamic component $h_D(t)$, the resultant channel envelope $|H(t)|$ follows the classical Rician distribution when $A \neq 0$ and reduces to Rayleigh distribution when $A = 0$. The probability density function (PDF) of the Rician-distributed envelope $|H|$ is:
\begin{equation}
	f_{|H|}(r)=\frac{r}{\sigma^2}\exp\left(-\frac{r^2+|A|^2}{2\sigma^2}\right)I_0\left(\frac{r|A|}{\sigma^2}\right), \quad r \geq 0
\end{equation}
where $I_0(\cdot)$ denotes the modified Bessel function of the first kind with order zero, and $\sigma^2$ is the variance of the dynamic Gaussian component. Therefore, it is important to note that we are not merging multiple static paths into a single physical path; rather, we treat their vector summation as a composite deterministic component. This approach aligns precisely with the fundamental assumption of the Rician fading model, which consists of a deterministic (specular) component combined with a diffuse scattering (random Gaussian) component.

\subsection{Model Assumptions and Limitations}
The key assumptions and current limitations underlying the RT-GSHCM should be proposed. A clear understanding of these factors is essential for a fair assessment of the model’s scope and for identifying directions for future refinement.
	\subsubsection{Dependency on measurement data}
	The channel parameters in proposed RT-GSHCM and the calibration of the stochastic component within 6GPCM both depend on measurement data. Channel measurement remains an important foundation for relevant research. Such measurement campaigns typically cover limited frequency bands and specific scenarios, constraining our model’s direct applicability to the scenarios and frequency ranges explicitly measured. Without corresponding measurement data, the model’s accuracy might degrade when generalized to other frequencies or significantly different propagation environments. 
	\subsubsection{High-Frequency Applicability}
 RT relies on geometrical optics (GO) and the geometrical theory of diffraction (GTD), which are known to be accurate with electrically smooth surfaces. However, above 24 GHz the dominance of surface roughness and diffuse scattering violates pure GO/GTD assumptions \cite{SmuldersMmwaveScattering}. Model accuracy may degrade unless higher‑order rough‑surface scattering modules or additional calibrations are employed.

\section{DCM Construction}
		\begin{figure*}[t]
	\centering
	\includegraphics[width=17cm]{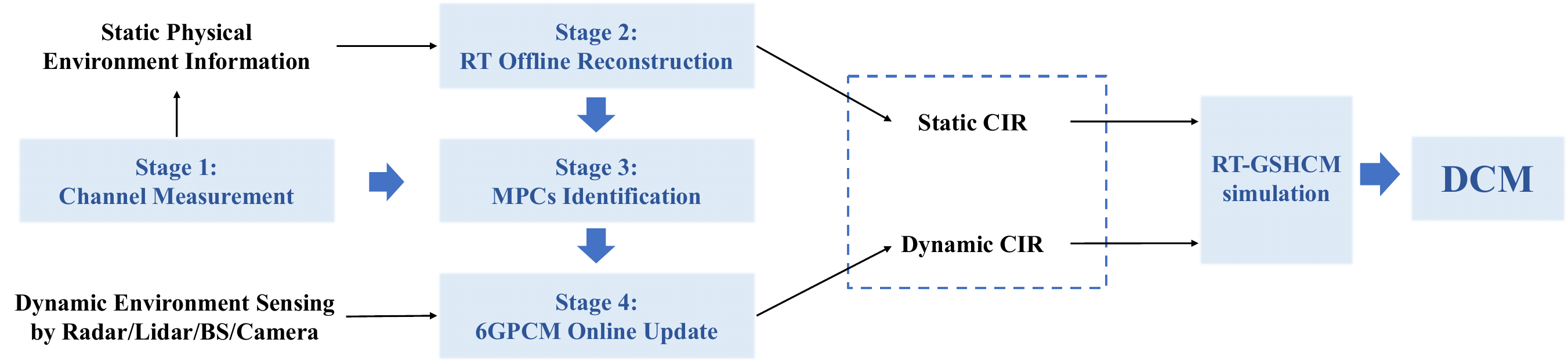}
	\caption{Construction procedure of the DCM.}
	\label{fig:HCM_procedure}
\end{figure*}
In this section, a sub-6 GHz channel measurement is conducted and the DCM for the measurement scenario is constructed. Specifically, we first model the static scattering channel by RT. Next, by comparing the MPCs of static scattering channel with estimated channel from measurement, the ratio between $K_{vu}(\ell)$ and $N_{vu}(t,\ell)$ in this scenario is obtained. Finally, Modeling the dynamic scattering channel by 6GPCM, we further construct the DCM for this area. Complete procedure of the DCM construction is illustrated in Fig. \ref{fig:HCM_procedure}.

	\subsection{Channel Measurement Campaign}
	
	\subsubsection{Measurement Environment and Setup}
		\begin{table}[b]
		\caption{System Setup.}
		\renewcommand{\arraystretch}{1.5}
		\begin{center}
			\begin{tabular}{|c|c|}
				\hline
				\textbf{Parameters} &  \textbf{Configurations} \\ \hline
				Carrier frequency (GHz)& 5.5\\ \hline
				Bandwidth (MHz) & 320\\ \hline
				Number of Rx antennas & 64\\ \hline
				Number of Tx antennas & 1\\ \hline
				Rx antenna array & \makecell{8×8 dual-polarization \\ cylindrical antenna array} \\ \hline
				Tx antenna array & Single omnidirectional antenna\\ \hline
				Tx height (m)& 30\\ \hline
				PN code length (chips) & 1023\\ \hline
				Transmit power (dBm)& 40\\ \hline
				Length of sounding signal (us)& 4\\ \hline
			\end{tabular}
			\label{tab:measurement parameter}
		\end{center}
	\end{table}
	
	The Keysight time-domain channel sounder was utilized for this measurement, with a distinct transmitter and receiver setup. The transmitter is equipped with an M8190A waveform generator, a power amplifier, a 32-channel switch, and synchronized by a GPS Rubidium clock. The receiver features an M9703B digital receiver, eight intermediate frequency (IF) amplifiers, an N5173B signal generator, and eight radio frequency (RF) switches, capable of 64-channel sequential measurements. This setup is designed for single-input multiple-output (SIMO) channel measurements. The received RF signal is mixed with a local oscillator to create the IF signal, which is then amplified by IF amplifiers and sampled at 1.6 GSa/s by the M9703B. The Keysight VSA 89600 software handles the downconversion, converting the data to baseband in-phase and quadrature format and storing it on a server. The system uses a single omnidirectional antenna for transmission and an 8×8 dual-polarization cylindrical antenna array for reception. Full system configurations are outlined in Table \ref{tab:measurement parameter}.
	
\begin{figure}[t]
		\centering
		\includegraphics[width=8.9cm]{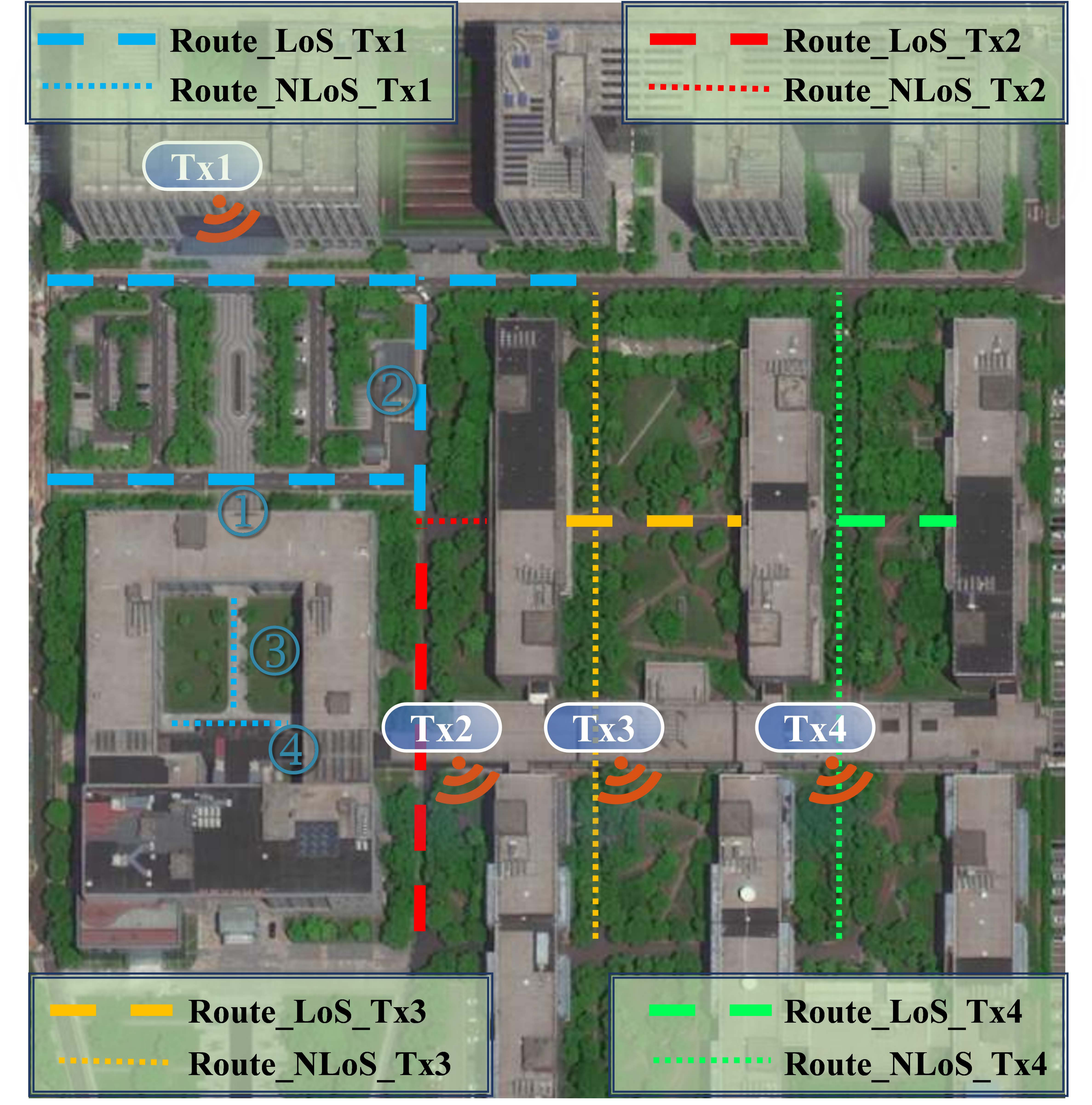}
		\caption{Overview of the measurement campaign.}
		\label{fig:measurement}
	\end{figure}

	The measurements were conducted at the ChinaNetwork Valley site in Nanjing, China, where four office buildings are about 37 meters high and Each building is separated by approximately 60 meters. The single omnidirectional antenna at the Tx side is fixed on a metal stand and placed on the eighth floor of an office building (named B1) at a height of 30 meters. The antenna array at the Rx side is attached to a trolley and moves on different routes within the venue. This measurement campaign includes both LoS and NLoS routes. NLoS propagation is mostly caused by the obstruction of buildings. The detailed measurement routes and their corresponding lengths are shown in Fig. \ref{fig:measurement}. Measurement locations are uniformly distributed across all routes. To ensure a sufficient number of valid samples, at least 30 snapshots are measured at each location on LoS and NLoS routes, respectively.
	
	\subsubsection{Measurement Data Process}
	Prior to conducting actual measurements, the channel sounder must be calibrated via a direct back-to-back connection to eliminate the system response inherent to the measurement setup. The method of data postprocessing is the same as \cite{III-1}. Here, we define the received signal obtained by direct connection calibration of system and the wireless received signal as $y_{cal}(t)$ and $y_{rec}(t)$, respectively. They can be expressed as 
	\begin{equation}
		y_{cal}(t) = x(t) \ast s(t) \label{y_cal}
	\end{equation}
	\begin{equation}
		y_{rec}(t) = x(t) \ast s(t) \ast h(t), \label{y_rec}
	\end{equation}
	where $x(t)$ denotes the transmitted signal and $s(t)$ denotes the back-to-back system response. $\{ \ast\}$ denotes the convolution operation. The CIR is denoted by $h(t)$. By the operation of FFT, $Y_{cal}(f)$ and $Y_{rec}(f)$ can be obtained, and the CIR can be further calculated by the inverse fast Fourier transform (IFFT)
	\begin{equation}
		h(t) = \mathbf{IFFT} \left\{ Y_{rec}(f) / Y_{cal}(f) \right\}.
	\end{equation}
	
	The space-alternating generalized expectation-maximization (SAGE) algorithm \cite{III-1} is used to extract MPCs from the delay PSD, which is calculated by averaging the data over all the snapshots by
	\begin{equation}
		P(\tau) = \left|  \frac{1}{S}\sum_{s=1}^{S}h_s(\tau)  \right|^2,
	\end{equation}
	where $S$ denotes the number of measured snapshots at each location. $h_s(\tau)$ denotes the CIR at the $s$th snapshot. $|\cdot|$ stands for absolute value. In the process of distinguishing between MPCs and background noise, a threshold is typically set at a level 6 dB above the noise floor, as referenced in \cite{III-2}. 
	In the SIMO static channel measurement, the estimated CIR can be characterized by \cite{III-3}
	\begin{equation}
		\hat{h}(t,\tau;\Theta_{l}) = \sum_{l=1}^L P_l \cdot \vec{c}_R(\boldsymbol{\Omega}_l^R)e^{j2\pi\mu_lt}\delta(\tau-\tau_l), \label{sage}
	\end{equation}
	where $L$ denotes the number of MPCs. The parameters' set of the $l$th MPC is denoted by $\Theta_{l}=[P_l,\boldsymbol{\Omega}_l^R,\tau_l]$, including the amplitude $P_l$, the direction of arrival $\boldsymbol{\Omega}_l^R$, and the propagation delay $\tau_l$. Here, $\boldsymbol{\Omega}_l^R$ can be described as an unit vector in spherical coordinates
	\begin{equation}
		\boldsymbol{\Omega}_l^R = [\sin(\theta_{l})\sin(\phi_{l}),\sin(\theta_{l})\cos(\phi_{l}),\cos(\theta_{l})],
	\end{equation}
	where $\theta_l$ and $\phi_l$ denote AAoA and EAoA, respectively. $\vec{c}_{R}(\boldsymbol{\Omega}_l^{R})$ in \eqref{sage} denotes the steering vector at Rx side. Using the high-resolution SAGE algorithm, the MPCs are then extracted from the measurements and will be compared with MPCs from RT simulation in the next subsection.
	
\subsection{Offline Pre-construction by RT}
\begin{figure}[t]
	\centering
	\subfloat[]
	{
		\hspace{-0.5cm}
		\includegraphics[width=5.6cm]{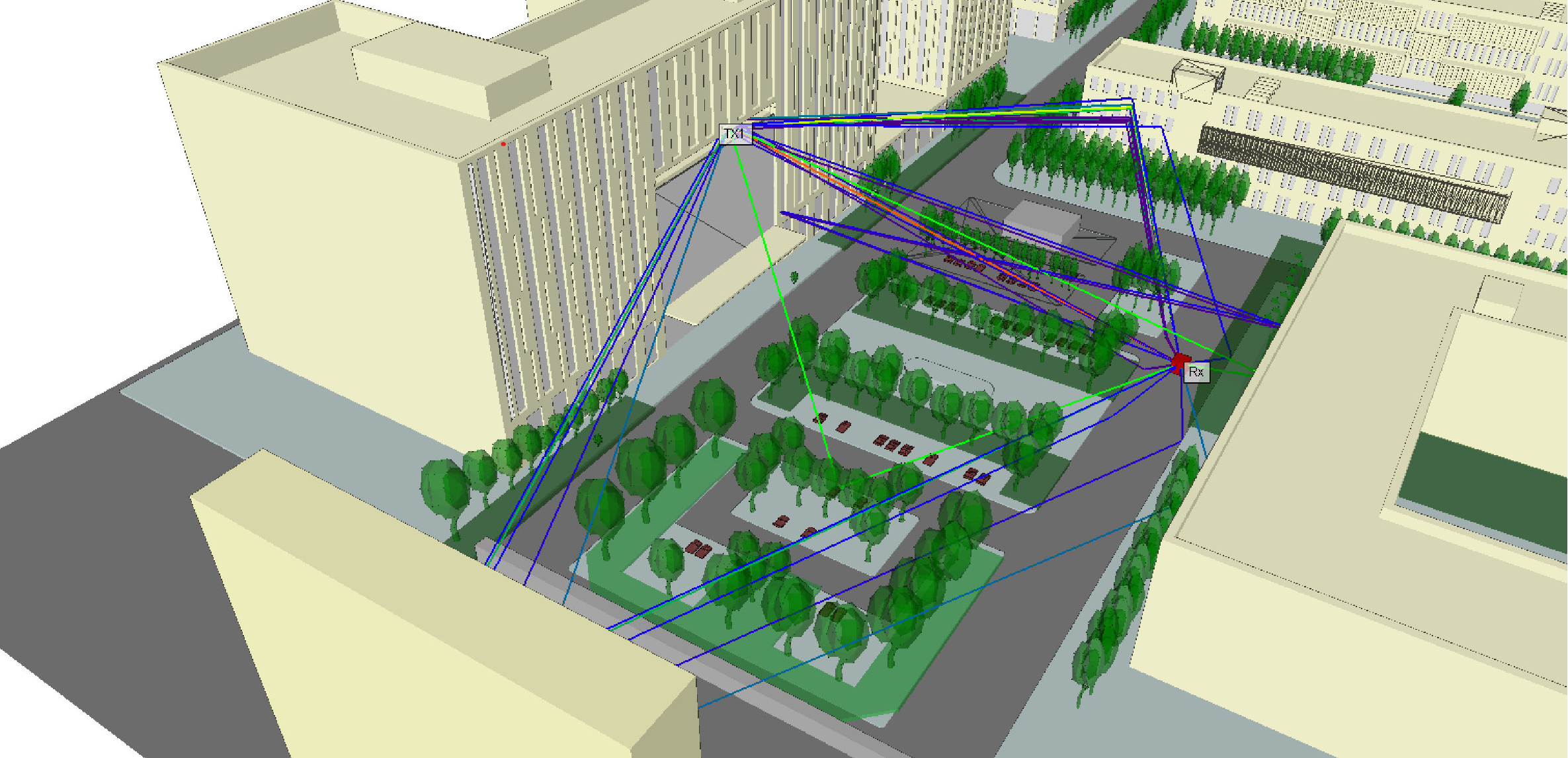}
		\label{fig:RT with path}
	}
	\subfloat[]
	{
	\includegraphics[width=3.4cm]{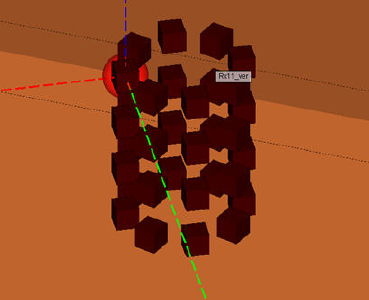}
	\label{fig:RT antenna}
	}\\
	\subfloat[]
	{
		\hspace{-1cm}
		\includegraphics[width=8.4cm]{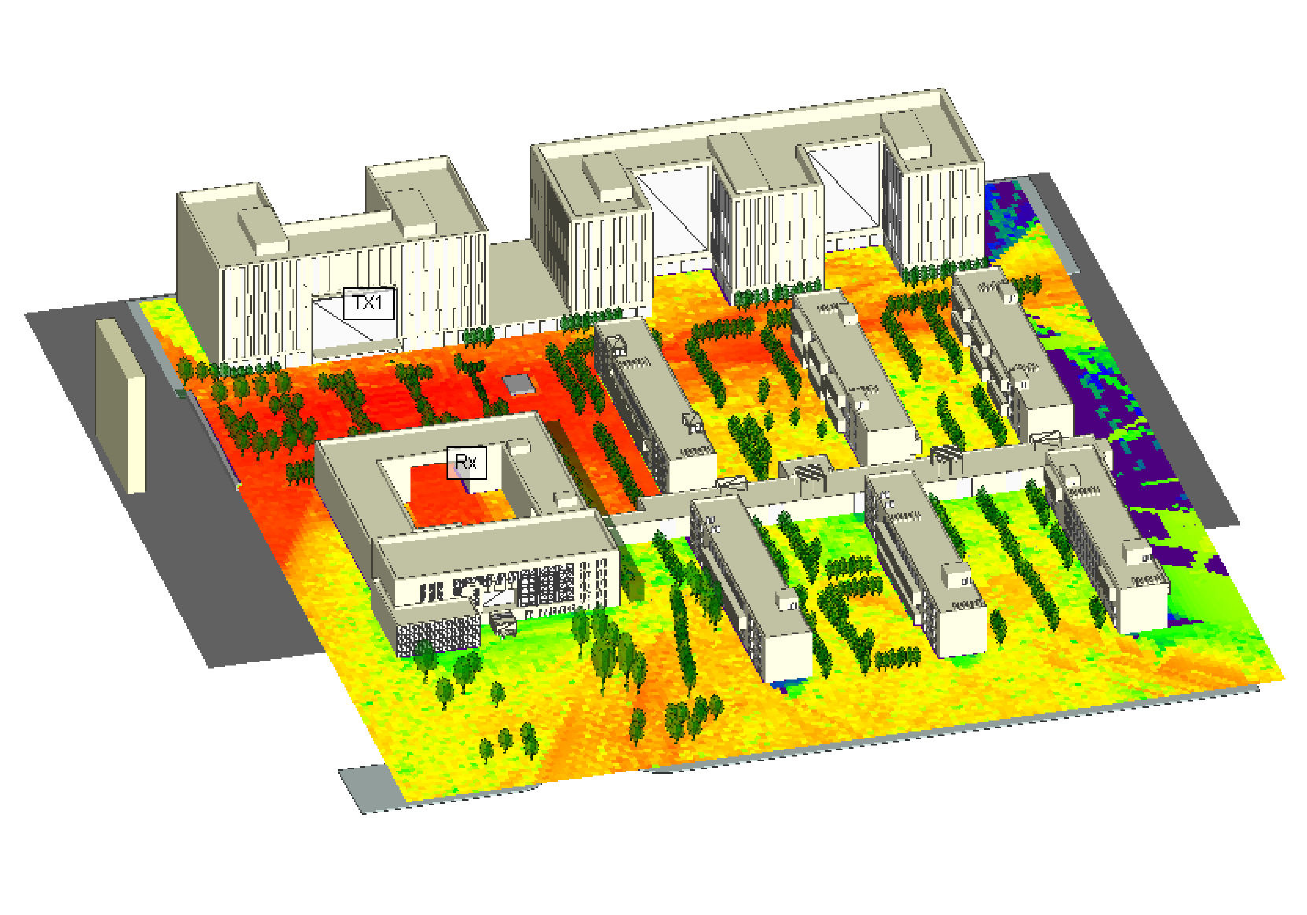}
		\label{fig:RT all scenario}
	}
	
	\caption{Communication environment reconstructed in WI. (a) Illustration of rays from RT. (b) Illustration of 8×8 dual-polarized cylindrical antenna array. (c) RSS distribution in measurement environment.}
	\label{fig:RT scenario}
\end{figure}
As shown in Fig. \ref{fig:RT scenario}, we reconstruct the scenario of ChinaNetwork Valley using Remcom Wireless Insite (WI), a wireless propagation software. The simulation configurations are the same as the measurement setup in Table \ref{tab:measurement parameter}. In this section, we consider various propagation mechanisms, including LoS and reflected signals. Fig. \ref{fig:RT scenario}\subref{fig:RT all scenario} illustrates some of these exemplary propagation paths. To maintain a manageable computational load for RT, we limit our analysis to reflections up to the sixth order. To obtain the parameters of static channels, only static IOs in environment including buildings and trees are setting in WI. By RT simulation, the ray parameters such as delays, angles, and powers are generated and further applied in \eqref{h^LoS} and \eqref{h^S} to implement the static component of the RT-GSHCM. 

It should be noted that WI does not support directly setting the radiation pattern for the entire MIMO antenna array. Thus, two co-located antenna elements with orthogonal polarizations (±45°) are used to simulate dual-polarized antennas. As shown in Fig.~\ref{fig:RT scenario}\subref{fig:RT antenna}, 32 dual-polarized antenna pairs were arranged as an intended 8×8 dual-polarized cylindrical array.

\subsection{MPCs Identification}

\begin{figure*}[t]
	\centering
	\subfloat[]{
		\includegraphics[width=9cm]{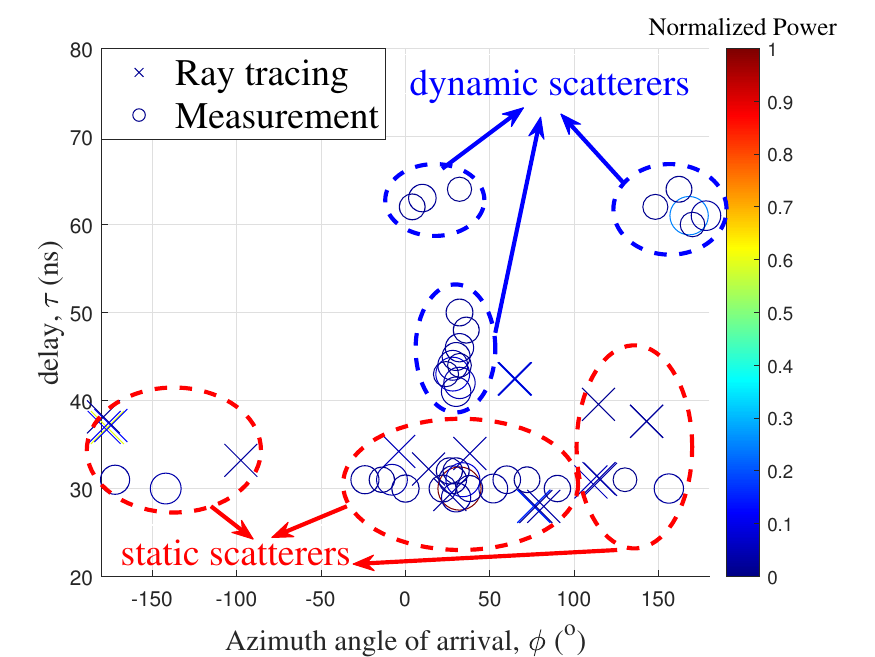}
		\label{fig:AoADPSD_pos1}
	}
	\subfloat[]{
		\includegraphics[width=9cm]{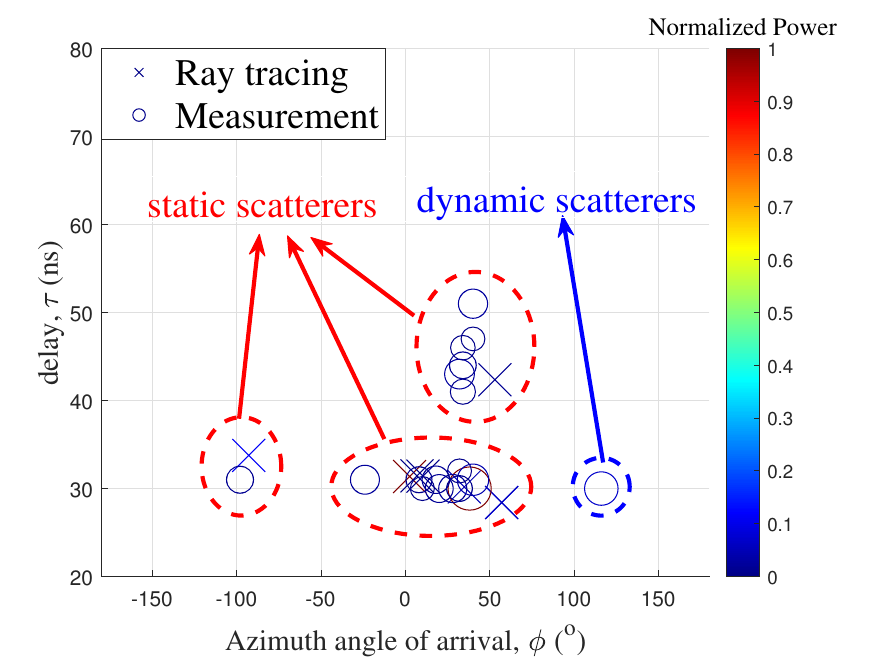}
		\label{fig:AoADPSD_pos2}
	}\\
	\subfloat[]{
		\includegraphics[width=9cm]{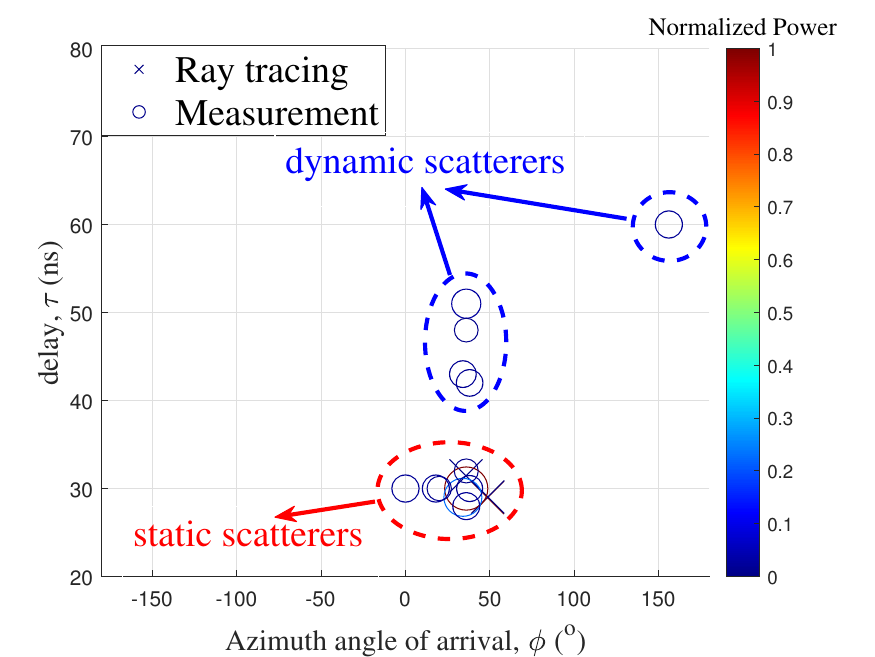}
		\label{fig:AoADPSD_pos3}
	}
	\subfloat[]{
		\includegraphics[width=9cm]{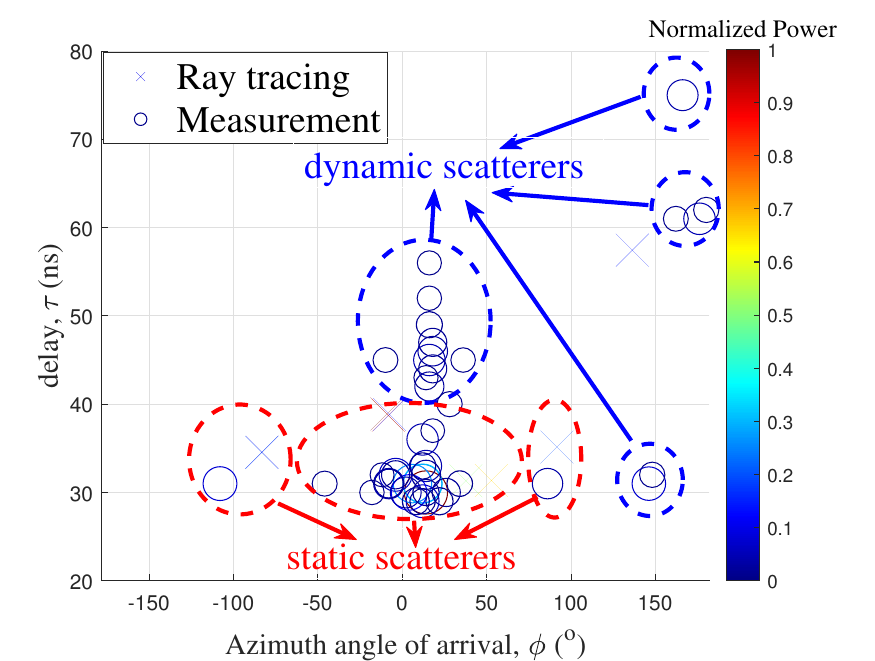}
		\label{fig:AoADPSD_pos4}
	}
	\caption{Normalized AAoA-delay PSDs of RT synthetic data and measurement results in Route\_LoS\_Tx1 at (a) location 3, (b) location 7, (c) location 9, and (d) location 13.}
	\label{fig:AAoADPSD}
\end{figure*}

Based on the parameters estimated by SAGE algorithm and the parameters of rays in RT simulation, the MPCs in RT channel (RT-MPCs) and MPCs in measurement channel (M-MPCs) can be illustrated in the joint AAoA and delay domain as shown in Fig. \ref{fig:AAoADPSD}, where each marker deontes one MPC. Symbol ‘$\times$’ denotes the RT-MPCs and symbol ‘$\bigcirc$’ denotes the M-MPCs. Colors of the symbols indicate the power of the corresponding MPCs. A RT-MPC are considered to be matched with a M-MPC if the Euclidean distance between them is smaller than a defined threshold. The matching MPCs are labeled with red dotted circles and other MPCs are labeled with pink solid circles, standing for the dynamic clusters. It can be found from Fig. \ref{fig:AAoADPSD} that each RT-MPC can be matched with at least one M-MPC and the power of matching MPCs is almost the same. However, there are still some observable M-MPCs, which cannot be matched with RT-MPCs. Because the dynamic IOs in physical environment are not considered in RT simulation, these M-MPCs may be caused by diffraction and diffuse scattering from dynamic IOs, corresponding to the dynamic clusters in the RT-GSHCM.

\begin{table}[t]
	\caption{Basic Simulation Settings.}
	\renewcommand{\arraystretch}{1.5}
	\begin{center}
		\begin{tabular}{|c|c|}
			\hline
			\textbf{Parameters} &  \textbf{Value} \\ \hline
			Simulation frequency (GHz) & 5.5\\ \hline
			Bandwidth (MHz) & 320\\ \hline
			Rx antenna array & \makecell{8×8 dual-polarization \\ cylindrical antenna array} \\ \hline
			Tx antenna array & Single omnidirectional antenna\\ \hline
			Simulation duration (s) & 1\\ \hline
			Number of clusters & 15\\ \hline
			Number of rays within one cluster & 10\\ \hline
			Sampling interval (ms) & 1\\ \hline
			Mobility of clusters (m/s) & 0.5\\ \hline
		\end{tabular}
		\label{tab:simulation parameter}
	\end{center}
\end{table}

\subsection{Online Update by 6GPCM}

By identifying the M-MPCs with different IOs in physical environment, we can obtain the value of $K_{vu}(\ell)$,  $N_{vu}(t,\ell)$, $K_S$, and $K_D$ in simulation setting of the proposed hybrid model, which can be further used in the accuracy evaluation to ensure a fair comparison. 
Other parameters in RT-GSHCM simulation setting are summarized in Table \ref{tab:simulation parameter}.

\section{Statistical Properties}
In this section, important statistical properties in space, time, frequency, angular, doppler, and delay domain are studied to verify the accuracy and correctness of the RT-GSHCM. Detailed information of the study of comprehensive statistical property can be referred to \cite{all_property}.

\subsection{Space-Time-Frequency Correlation Function (STFCF)}
The STFCF can be calculated as \cite{all_property}
\begin{equation}
	\begin{aligned}
		&R_{vu,\tilde{v}\tilde{u}}\left(t,f,\ell;\Delta r^T,\Delta r^R,\Delta t,\Delta f,\Delta \ell \right) \\ &=\operatorname{E}\left[H_{vu}\left(t,f,\ell\right)H_{\tilde{v}\tilde{u}}^{*}\left(t+\Delta t,f+\Delta f,\ell+\Delta \ell \right)\right], \label{STFCF}
	\end{aligned}
\end{equation}
where $\operatorname{E}$ denotes the statistical expectation operation, $(\cdot)^{*}$ stands for the complex conjugation operation. $\Delta r^{T/R}$ denotes the antenna element spacing between $A^{T/R}_{v/u}$ and $A^{T/R}_{\tilde{v}/\tilde{u}}$, as $\Delta r^T = \delta_{\tilde{v}} - \delta_v$, and $\Delta r^R = \delta_{\tilde{u}} - \delta_u $. By substituting \eqref{CTF} into \eqref{STFCF}, the STFCF can be rewritten as
\begin{equation}
	\begin{aligned}
		&R_{vu,\tilde{v}\tilde{u}}\left(t,f,\ell;\Delta r^T,\Delta r^R,\Delta t,\Delta f,\Delta \ell\right) \\
		&=\frac{1}{K_S^{-1}+K_D^{-1}+1}R^{L}_{vu,\tilde{v}\tilde{u}}\left(f,\ell;\Delta r^T,\Delta r^R,\Delta f,\Delta \ell\right) \\ 
		&+\frac{K_S^{-1}}{K_S^{-1}+K_D^{-1}+1}R^{S}_{vu,\tilde{v}\tilde{u}}\left(f,\ell;\Delta r^T,\Delta r^R,\Delta f,\Delta \ell\right)\\
		&+\frac{K_D^{-1}}{K_S^{-1}+K_D^{-1}+1}R^{D}_{vu,\tilde{v}\tilde{u}}\left(t,f,\ell;\Delta r^T,\Delta r^R,\Delta t,\Delta f,\Delta \ell\right). 
	\end{aligned}
\end{equation}

Here, $R^{L}_{vu,\tilde{v}\tilde{u}}\left(f,\ell;\Delta r^T,\Delta r^R,\Delta f,\Delta \ell\right)$ denotes the STFCF of LoS component in static scattering channel
\begin{equation}
	\begin{aligned}
		&R^{L}_{vu,\tilde{v}\tilde{u}}\left(f,\ell;\Delta r^T,\Delta r^R,\Delta f,\Delta \ell \right) \\
		&=E\left\{H_{vu}^{L}\left(f,\ell\right)H_{\tilde{v}\tilde{u}}^{L *}\left(f+\Delta f, \ell + \Delta \ell\right)\right\} \\
		&=e^{j2\pi\left((f_c-f)\left[\tau_{vu}^L(\ell)-\tau_{\tilde{v}\tilde{u}}^L(\Delta \ell)\right] - \Delta f \tau_{vu}^L(\Delta \ell) \right)}.
	\end{aligned}
\end{equation}

Also, $R^{S}_{vu,\tilde{v}\tilde{u}}\left(f,\ell;\Delta r^T,\Delta r^R,\Delta f,\Delta \ell\right)$ in \eqref{STFCF} is the STFCF of NLoS component in static scattering channel. It can be calculated as
\begin{equation}
	\begin{aligned}
		&R^{S}_{vu,\tilde{v}\tilde{u}}\left(f,\ell;\Delta r^T,\Delta r^R,\Delta f,\Delta \ell\right) \\
		&=E\left\{H_{vu}^{S}\left(t,f\right)H_{\tilde{v}\tilde{u}}^{S *}\left(t+\Delta t,f+\Delta f\right)\right\} \\
		&=E \left\{\sum_{k=1}^{K_{vu}(\ell)} \sum_{k'=1}^{K_{\tilde{v}\tilde{u}}(\ell + \Delta \ell)} \sqrt{P_{vu,k}(\ell) \cdot P_{\tilde{v}\tilde{u},k'}(\ell+\Delta \ell)}\right.  \\
		&\left. \cdot e^{j2\pi\left((f_c-f) [\tau_{vu,k}(\ell) - \tau_{\tilde{v}\tilde{u},k'}(\ell+\Delta \ell)]-\Delta f \tau_{\tilde{v}\tilde{u},k'}(\ell+\Delta \ell) \right)} \right\}.
	\end{aligned}
\end{equation}

Finally, the STFCF of dynamic scattering channel is
\begin{equation}
	\begin{aligned}
		&R^{D}_{vu,\tilde{v}\tilde{u}}\left(t,f,\ell;\Delta r^T,\Delta r^R,\Delta t,\Delta f,\Delta \ell\right) \\
		&=E\left\{H_{vu}^{D}\left(t,f,\ell\right)H_{\tilde{v}\tilde{u}}^{D *}\left(t+\Delta t,f+\Delta f,\ell+\Delta \ell\right)\right\} \\
		&=E \left\{ \sum_{n=1}^{N_{vu}(t,\ell)} \sum_{m=1}^{M_n} \sum_{n'=1}^{N_{\tilde{v}\tilde{u}}(t+\Delta t,\ell+ \Delta \ell)}  \sum_{m'=1}^{M_{n'}} \sqrt{P_{vu,m_n}(t,\ell)} \right. \\
		& \cdot \sqrt{P_{\tilde{v}\tilde{u},m'_n}(t+\Delta t,\ell+\Delta \ell)} \cdot e^{-j2\pi \frac{\Delta f}{\lambda f_c} \cdot d^T_{\tilde{v},m'_n}(t+\Delta t,\ell+\Delta \ell)}\\
		& \cdot e^{-j2\pi \frac{\Delta f}{\lambda f_c} \cdot d^R_{\tilde{u},m'_n}(t+\Delta t,\ell+\Delta \ell)} \cdot e^{j2\pi \frac{(f_c-f)}{\lambda f_c} \left[d^T_{v,m_n}(t,\ell)+d^R_{u,m_n}(t,\ell)\right] }\\
		&\left. \cdot e^{-j2\pi \frac{(f_c-f)}{\lambda f_c} \left[d^T_{\tilde{v},m'_n}(t+\Delta t,\ell+\Delta \ell)+d^R_{\tilde{u},m'_n}(t+\Delta t,\ell+\Delta \ell)\right] }  \right\}.
	\end{aligned}
\end{equation}

The STFCF can be decomposed into spatial cross-correlation function (CCF) of antenna array, geographical CCF in channel map, temporal autocorrelation function (TACF), and FCF by setting \{$\Delta t = 0, \Delta f = 0, \Delta \ell = 0$\}, \{$\Delta r^{T/R} = 0, \Delta t = 0, \Delta f = 0$\}, \{$\Delta r^{T/R} = 0, \Delta f = 0, \Delta \ell = 0$\}, and\{$\Delta r^{T/R} = 0, \Delta t = 0, \Delta \ell = 0$\}. 
\vspace{-1em}

\subsection{RMS Angular Spread and LCR}

The RMS angular spread at Rx side is the square root of the second-order central moment of angular PSD, defined as
\begin{equation}
	\begin{aligned}
		&\sigma^{\mathrm{AoA}}_{vu,\tilde{u}}(t,f,\ell;\phi_{\mathrm{AoA}}) \\
		&=\sqrt{\frac{\mathop{\int}\limits_{0}^{2\pi}(\phi_{\mathrm{AoA}}-u_{\mathrm{AoA},vu,\tilde{u}}\left(t,f,\ell\right))^{2}D_{vu,\tilde{u}}\left(t,f,\ell;\phi_{\mathrm{AoA}}\right)d\phi_{\mathrm{AoA}}}{\int_{0}^{2\pi}D_{vu,\tilde{u}}\left(t,f,\ell;\phi_{\mathrm{AoA}}\right)d\phi_{\mathrm{AoA}}}},
	\end{aligned}
\end{equation}
where $u_{\mathrm{AoA},vu,\tilde{u}}\left(t,f,\ell\right)$ is the average AoA. The angular PSD is denoted as $D_{vu,\tilde{u}}(t,f,\ell;\phi_{AoA})$, which can be obtained from the spatial-Doppler PSD 	$D_{vu,\tilde{u}}\left(t,f,\ell;\varpi^R\right)$, defined as the Fourier transform of spatial CCF at Rx side. The transformer is complicated and the details can be refered to \cite{all_property}. Here, $\varpi$ is the spatial-Doppler frequency variable. We define the angle between the AoA wave and the orientation of Rx antenna array as $\theta^R$, then $\varpi^R = \cos{\theta^R}$ \cite{second_order}.

By counting the times the envelope crosses different threshold levels, the LCR at specific level R is expressed as \cite{Patzold1998}
\begin{equation}
	\begin{aligned}
		L(R)& =\quad\frac{2R\sqrt{K+1}}{\pi^{3/2}}\sqrt{\frac{b_{2}}{b_{0}}-\frac{b_{1}^{2}}{b_{0}^{2}}} e^{-K-(K+1)R^{2}} \\
		&\times\quad\int_0^{\pi/2}\cosh\left(2\sqrt{K(K+1)}R\cos\theta\right) \\
		&\times\quad\left[e^{-(\chi\sin\theta)^{2}}+\sqrt{\pi}\chi\sin\theta \mathrm{erf}(\chi\sin\theta)\right]d\theta,
	\end{aligned}
\end{equation}
where $\mathrm{erf}(\cdot)$ is the error function,
\begin{equation}
	\chi = \sqrt{Kb_1^2/(b_0b_2-b_1^2)}
\end{equation}
\begin{equation}
	b_i = \frac{K}{j^i  K_S} \cdot \frac{d^i R^S_{vu,\tilde{u}}(\Delta r^R)}{d(\Delta r^R)^i}+\frac{K}{j^i  K_D} \cdot \frac{d^i R^D_{vu,\tilde{u}}(\Delta r^R)}{d(\Delta r^R)^i}.
\end{equation}

\subsection{RMS Doppler Spread}
The motion of dynamic scatterers cases the channel dispersion in Doppler frequency domain. RMS Doppler spread is often used to measure these dispersions,  given by
\begin{equation}
	\sigma_{\xi,vu}\left(t,f\right)=\sqrt{\frac{\int\limits_{0}^{\infty}(\xi-u_{\xi,vu}\left(t,f,\ell\right))^{2}D_{vu}\left(t,f,\ell;\xi\right)d\xi}{\int\limits_{0}^{\infty}D_{vu}\left(t,f,\ell;\xi\right)d\xi}},
\end{equation}
where $D_{vu}(t,f,\ell;\xi)$ denotes Doppler PSD and $u_{\xi,vu}\left(t,f,\ell\right)$ is the average Doppler shift. Here, $\xi$ is Doppler frequency. Similarly to angular PSD, Doppler PSD is the Fourier transformer of TACF $R_{vu}\left(t,f,\ell;\Delta t\right)$ w.r.t. the time interval $\Delta t$.

\subsection{FCF, Delay PSD, and RMS Delay Spread}
FCF measures the frequency correlation of the channel as
\begin{equation}
	\begin{aligned}
		&R_{vu}\left(t,f,\ell;\Delta f\right)\\
		&= \frac{K_S^{-1}}{K_S^{-1}+K_D^{-1}+1} \cdot E\left\{\sum_{k=1}^{K_{vu}(\ell)}P_{vu,k}(\ell)\cdot e^{-j2\pi\Delta f\tau_{vu,k}(\ell)}\right\}\\
		& + \frac{1}{K_S^{-1}+K_D^{-1}+1} \cdot e^{-j2\pi\Delta f\tau_{vu}^L(\ell)} + \frac{K_D^{-1}}{K_S^{-1}+K_D^{-1}+1} \\
		& \cdot E \left\{\sum_{n=1}^{N_{vu}(t,\ell)}\sum_{m=1}^{M_{n}}P_{vu,m_{n}}(t,\ell)\cdot e^{-j2\pi\Delta f \frac{ \left[d_{v,m_{n}}^{T}(t,\ell)+d_{u,m_{n}}^{R}(t,\ell)\right]}{\lambda f_{c}}} \right\} \label{FCF}.
	\end{aligned}
\end{equation}

$D_{vu}(t,f,\ell;\tau)$ denotes delay PSD, which is the inverse Fourier transform of FCF w.r.t. $\Delta f$, obtained as \cite{all_property}
\begin{equation}
	D_{vu}\left(t,f,\ell;\tau\right)=\int R_{vu}\left(t,f,\ell;\Delta f\right)e^{j2\pi\tau\Delta f}d\Delta f.
\end{equation}

Similar to the RMS angular spread and RMS Doppler spread, the RMS delay spread is defined as the square root of the second-order central moment of the delay PSD. It can be calculated as
\begin{equation}
	\sigma_{\tau,vu}\left(t,f,\ell\right)=\sqrt{\frac{\int\limits_{0}^{\infty}(\tau-u_{\tau,vu}\left(t,f,\ell\right))^{2}D_{vu}\left(t,f,\ell;\tau\right)d\tau}{\int\limits_{0}^{\infty}D_{vu}\left(t,f,\ell;\tau\right)d\tau}} \label{DS},
\end{equation}
where $u_{\tau,vu}\left(t,f,\ell\right))$ is the average delay.

	\section{Evaluation and Analysis}
	In this section, the accuracy and time-efficiency of DCM are validated. To explore the correctness of RT-GSHCM and its scalability in other scenarios, the channel properties in space, time, and frequency domain are simulated and analyzed under various cluster configurations.
	\vspace{-0.5em}
	\subsection{Accuracy Evaluation}
\begin{figure*}[t]
		\centering
		\subfloat[]
		{
			\includegraphics[width=6cm]{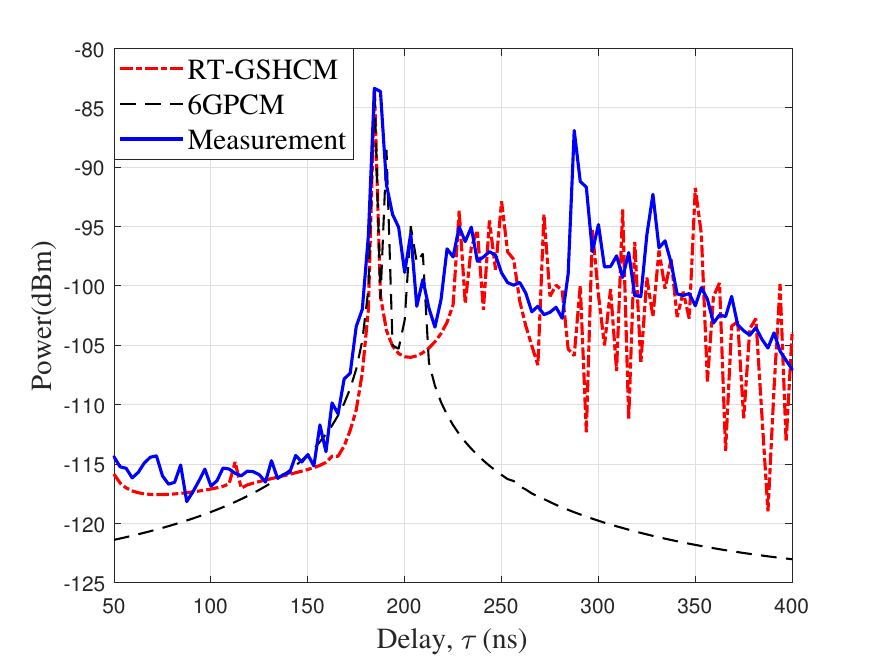}

		}
		\subfloat[]
		{
			\includegraphics[width=6cm]{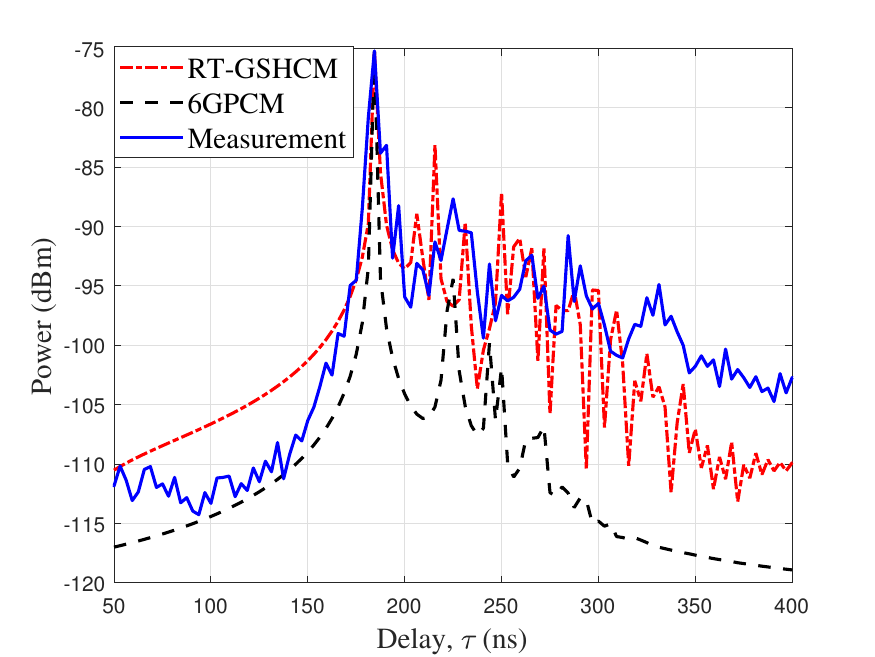}

		}
		\subfloat[]
		{
			\includegraphics[width=6cm]{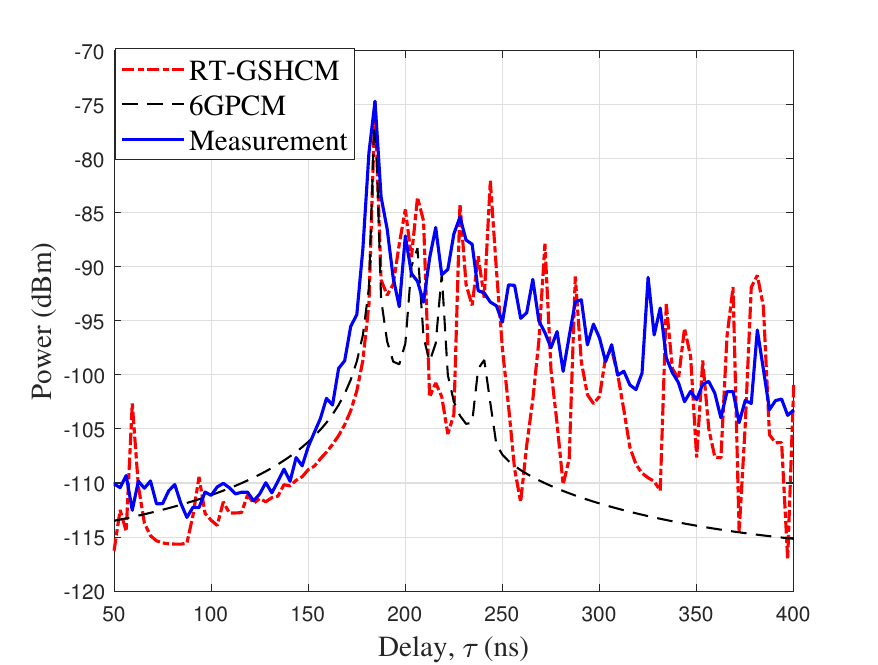}

		}	
		\caption{Delay PSDs of the data generated by RT-GSHCM, measurements, and 6GPCM in Route\_LoS\_Tx1 at (a) location 7, (b) location 17, (c) location 19.}
		\label{fig:delayPSD_Tx1}
	\end{figure*}

\begin{figure*}[t]
	\centering
	\subfloat[]
	{
		\includegraphics[width=6cm]{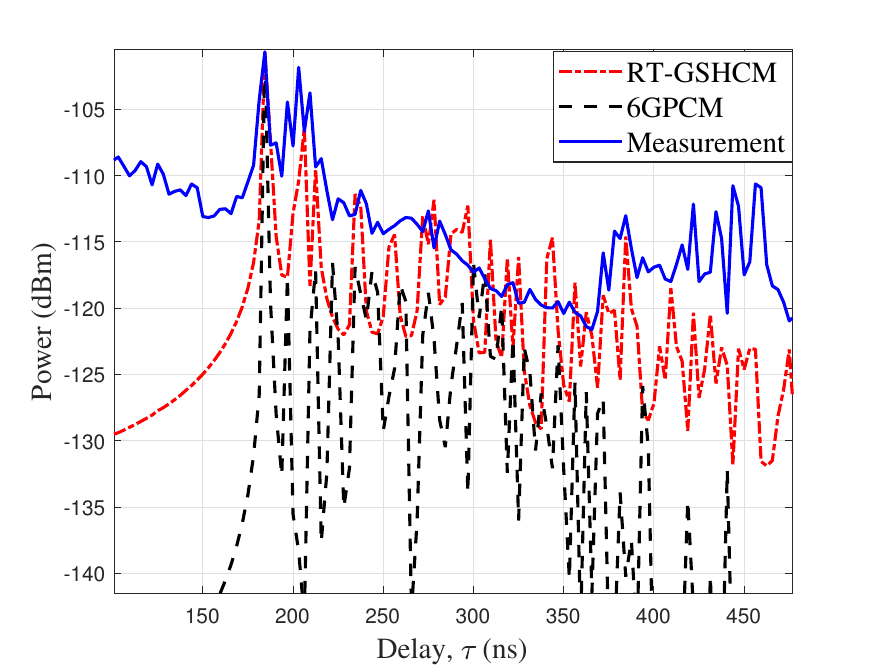}
	}
	\subfloat[]
	{
		\includegraphics[width=6cm]{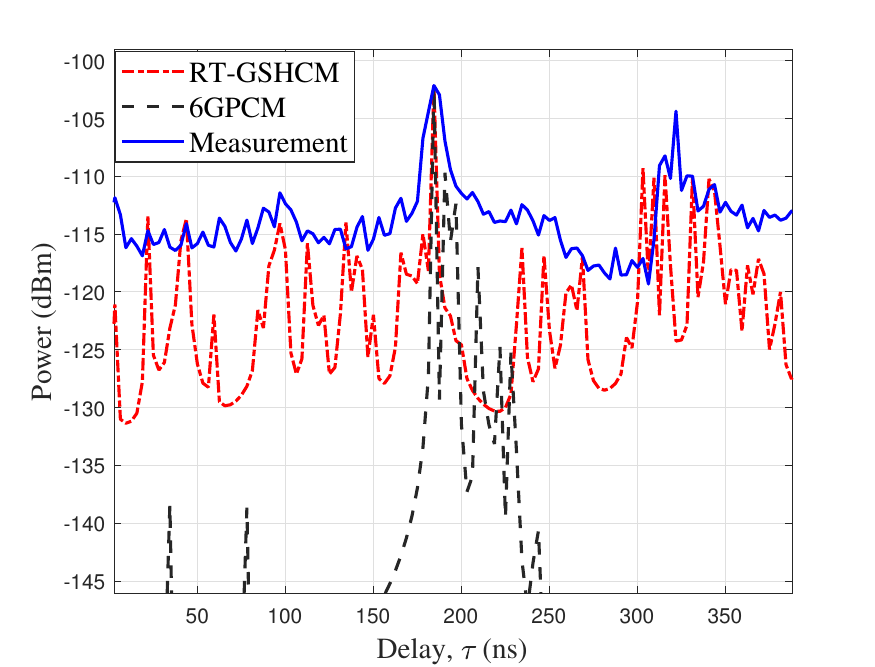}
	}
	\subfloat[]
	{
		\includegraphics[width=6cm]{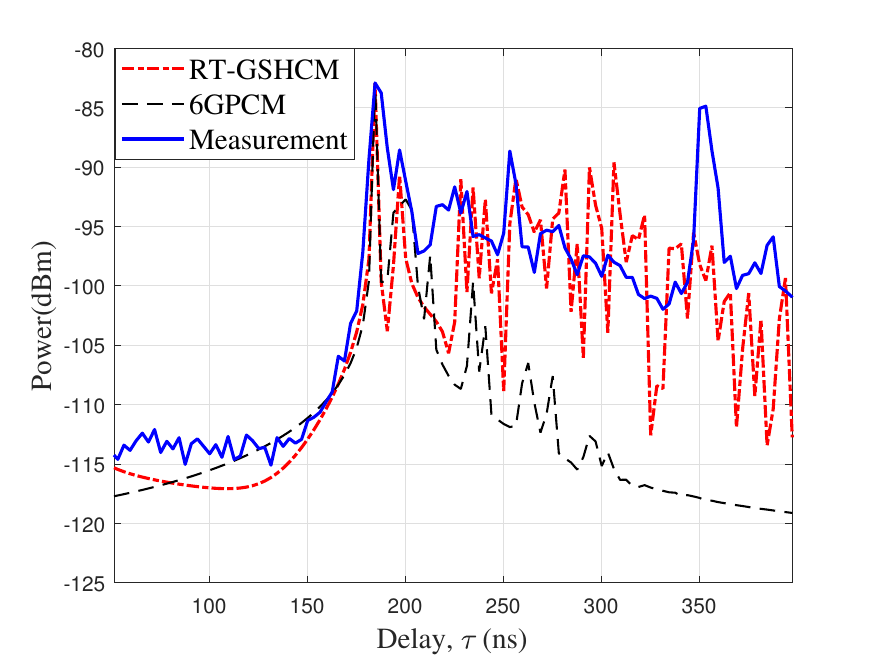}
	}	
	\caption{Delay PSDs of the data generated by RT-GSHCM, measurements, and 6GPCM in Route\_NLoS\_Tx3 at (a) location 6, (b) location 8, (c) location 22.}
	\label{fig:delayPSD_Tx3}
\end{figure*}

	Given the inherent variability of wireless channels, a direct comparison between synthesized CIR and actual measurement data is not comprehensive for model evaluation. Instead, an alternative approach is to assess the model's validity by examining the statistical properties of the wireless channels in a stochastic manner. In this work, the accuracy of DCM is verified by validating the effectiveness of the RT-GSHCM. 
	The comparison between the delay PSD of RT-GSHCM, measurement, and 6GPCM in Route\_LoS\_Tx1 is shown in Fig.~\ref{fig:delayPSD_Tx1}, We can see that the RT-GSHCM accord well with measurement than 6GPCM in specific positions. The reason is that the proposed RT-GSHCM contains accurate MPCs information from RT simulation. Additionally, most measurement positions in Route\_NLoS\_Tx3 are obscured by trees as shown in Fig.~\ref{fig:measurement}. Therefore, the same comparison in this scenario is conducted in Fig.~\ref{fig:delayPSD_Tx3} to demonstrate the generalization of RT-GSHCM. Fig.~\ref{fig:delayPSD_Tx3} demonstrates that RT-GSHCM maintains excellent agreement with measurement data. The close agreement in all cases confirms that RT-GSHCM accurately captures scenario-dependent channel behavior and thus generalizes robustly to varied propagation environments.

\begin{figure}[t]
	\centering
	\subfloat[]
	{
		\hspace{-5mm}
		\includegraphics[width=4.8cm]{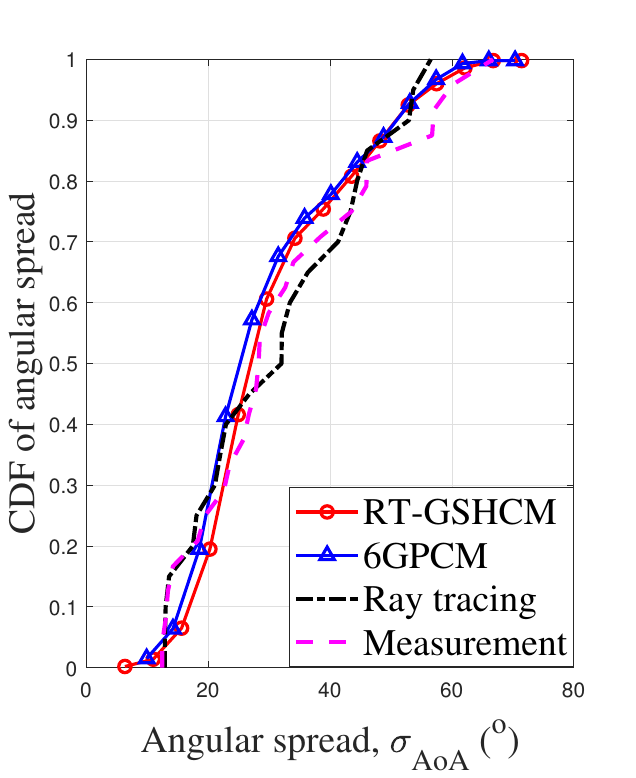}
		\label{fig:CDF of AS}
	}
	\subfloat[]
	{
		\hspace{-5mm}
		\includegraphics[width=4.8cm]{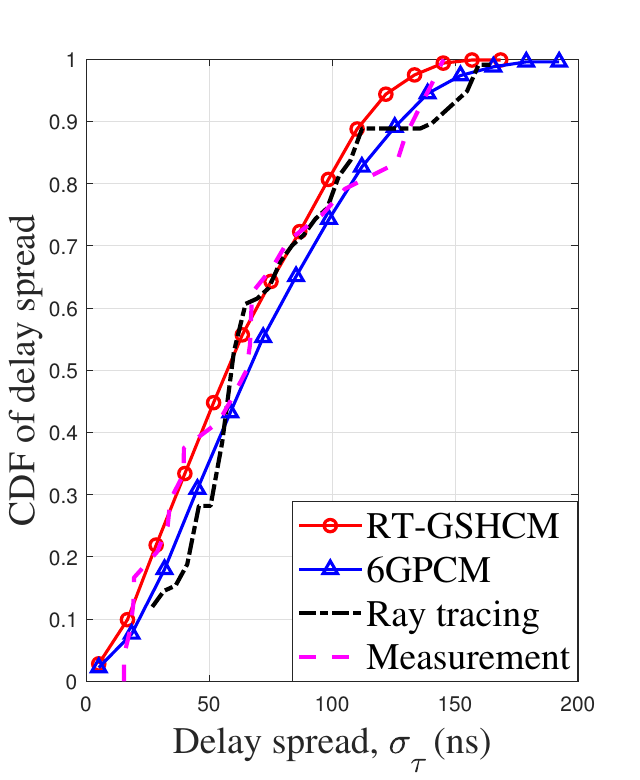}
		\label{fig:CDF of DS}
	}
	\caption{CDFs of (a) RMS angular spread and (b) RMS delay spread of RT-GSHCM, measurement results, RT, and 6GPCM.}
	\label{fig:CDF of AS&DS}
\end{figure}

	 In additional, more classic statistical properties are also explored on Route\_LoS\_Tx1. In Fig.~\ref{fig:CDF of AS&DS}, the CDFs of RMS angular spread and RMS delay spread of RT-GSHCM, channel measurement, RT, and 6GPCM are compared. It illustrates that the RMS angular spread and RMS delay spread of RT-GSHCM, RT, and 6GPCM all accord well with the measurement results.

\begin{figure}[t]
	\centering
	\includegraphics[width=9.6cm]{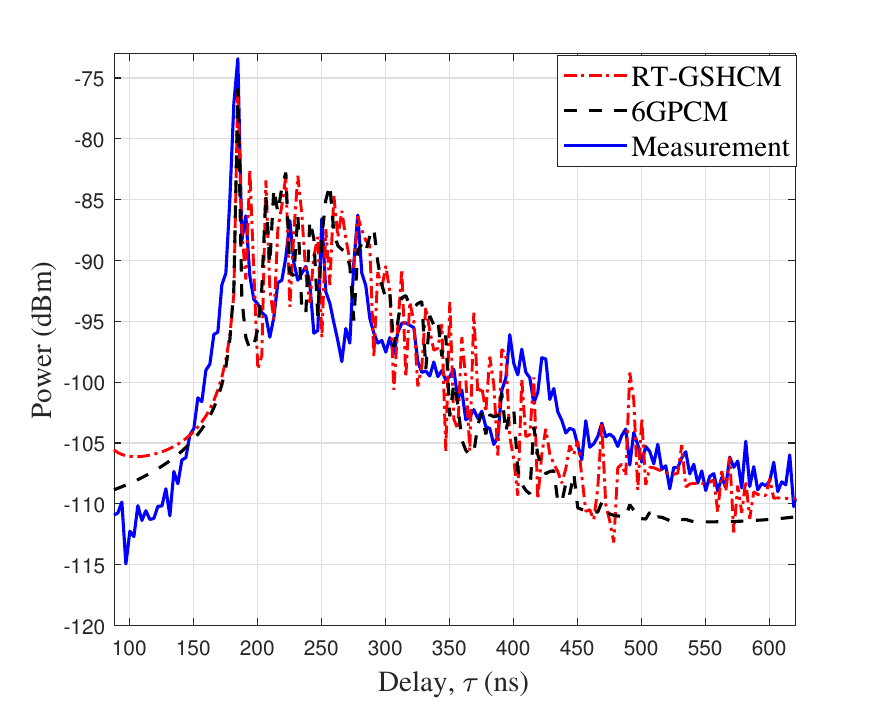}
	\caption{Average delay PSDs of the data generated by RT-GSHCM, measurement results, and 6GPCM in all locations of Route\_LoS\_Tx1.}
	\label{fig:DelayPSD_all}
\end{figure}

\vspace{-0.5em}
	\subsection{Time-Efficiency}
		\begin{table}[b]
		\caption{Computation Time for channel map updating.}
		\renewcommand{\arraystretch}{1.5}
		\begin{center}
			\begin{tabular}{|c|c|c|}
				\hline
				\textbf{Channel maps} &  \textbf{Construction methods} &  \textbf{Time cost on updating (s)} \\ \hline
				CKM & RT & 874\\ \hline
				REM & Channel measurements & Infinity\\ \hline
				DCM & RT-GSHCM & 0.4\\ \hline
			\end{tabular}
			\label{tab:updatetime}
		\end{center}
	\end{table}
	Channel maps need to be updated as the physical environment changes. Note that the channel map initial construction allows for substantial time consumption, but subsequent updates must be highly time-efficient to accommodate the needs of 6G MIMO communication systems. In order to assess the time-efficiency of DCM, we compare the time cost on updating between CKM, REM, and DCM. This comparison is made within the context of the scenario depicted in Fig. \ref{fig:measurement}, utilizing a system equipped with an Intel Core i9 processor 14900 and 32 GB of RAM. The RT setting is the same as Table \ref{tab:simulation parameter}. As the comparison results shown in Table \ref{tab:updatetime}, the CKM constructed by RT consumes 874 seconds to track and calculate all the parameters of 120 rays while the DCM only takes 0.4 seconds to update the channel information. The reason is that the CKM is constructed offline, the map update means reconstrution, which leads to the high time cost. However, Although the DCM also ultilize RT to implement the offline construction, it supports online updating with 6GPCM, which ensures the time-efficiency of DCM. 

	\subsection{Channel Properties and Analysis}
	To study the channel statistical properties of the RT-GSHCM, we compare the channel statistical properties with different cluster configurations. The main parameters of the cluster in the RT-GSHCM are carrier frequency, the number of dynamic clusters, and the speed of dynamic clusters, which are denoted as $\mathrm{f}$, $\mathrm{N}_{cluster}$, and $\mathrm{v}_{cluster}$, respectively.
	
	\subsubsection{RMS Angular Spread}
		\begin{figure}[t]
	\centering
	\includegraphics[width=9.2cm]{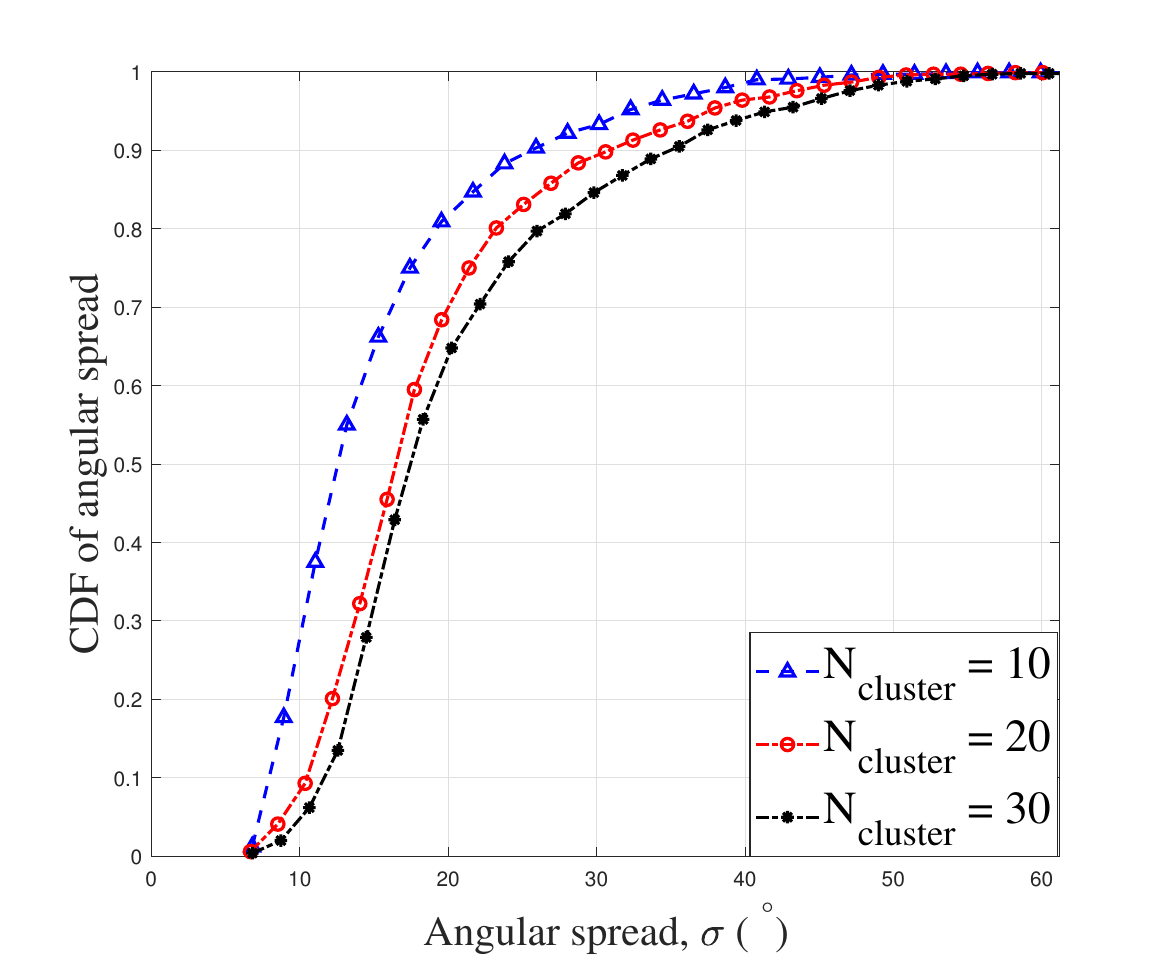}
	\caption{CDFs of RMS angular spread with different dynamic cluster numbers.}
	\label{fig:AS_clusternumber}
\end{figure}
	CDFs of angular spread of the synthetic data from RT-GSHCM with different dynamic cluster numbers are simulated as shown in Fig. \ref{fig:AS_clusternumber}. The increase of dynamic cluster number enhance the large angle MPCs' power. The reason is that the scattering from dynamic IOs always provides MPCs with large angle due to the motion of clusters.
	
	Another interesting discovery is that the maximum and minimum values of angular spread remain identical across different dynamic cluster numbers. This is attributed to the fact that, the range of angular spread for dynamic scattering channels is set as a subset of that for static scattering channels during simulation, thereby facilitating the differentiation of dynamic MPCs from the total channel. Consequently, it is the static scattering channels that determine both the upper and lower bounds of angular spread.

	\begin{figure}[b]
	\centering
	\includegraphics[width=9.2cm]{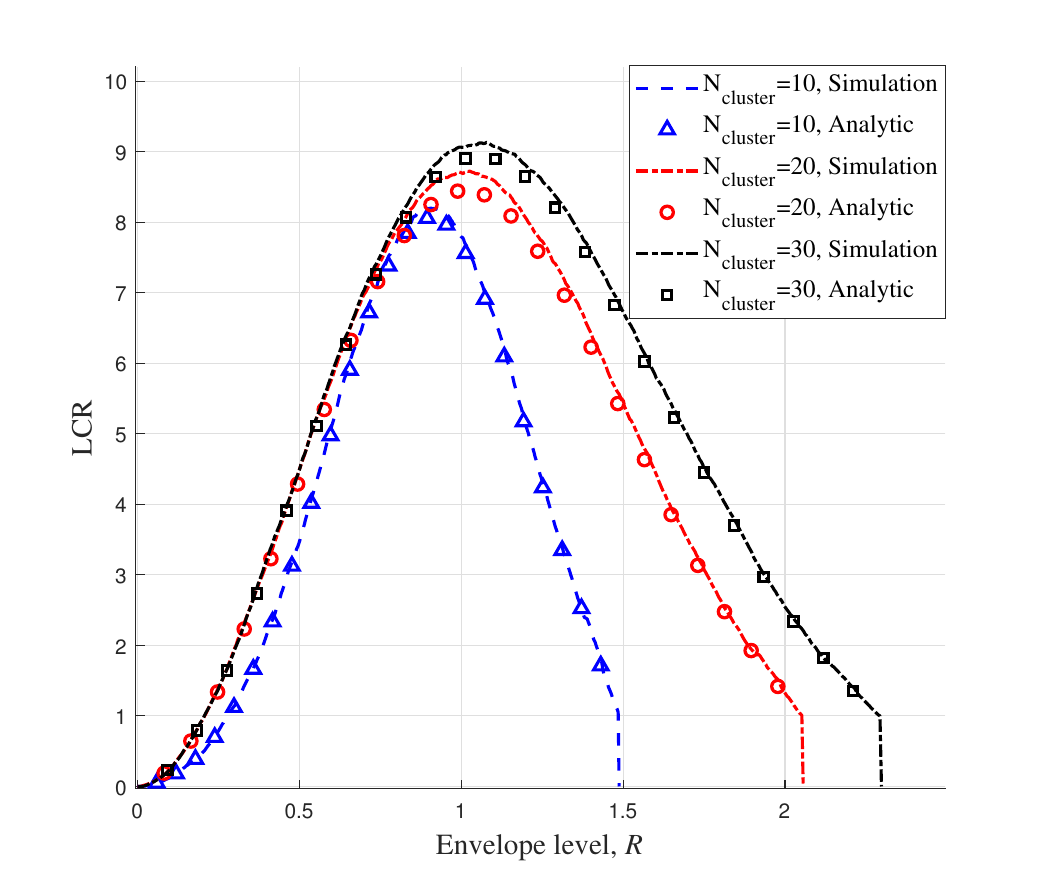}
	\caption{LCRs with different dynamic cluster numbers.}
	\label{fig:LCR_clusternumber}
\end{figure}

\subsubsection{LCR}
Fig. \ref{fig:LCR_clusternumber} shows the LCRs of RT-GSHCM with different dynamic cluster numbers. When the envelope level is less than the peak value of the curve, changes in the dynamic cluster number have a minimal impact on LCR. However, when the envelope level exceeds the peak value, the LCR increases with more dynamic clusters under the same threshold. The reason is that the signal envelop in low level is determined by the static scattering components in signal, which is a constant. The maximum LCR also increases with the increase of dynamic cluster number. This is mainly because more dynamic IOs lead to intenser changes in the physical environment. Simulation results accord well with the real change in physical environment, illustrating the correctness of the RT-GSHCM.

Moreover, it can be observed that the simulated LCR values are systematically slightly higher than the analytical results, particularly around the peak region, which aligns with observations from previous studies. This small bias is well documented in \cite{Patzold1998}. The analytical formula assumes an infinite scatterer field and an ideal continuous Doppler/space-correlation structure, whereas our finite-sinusoid simulator uses a limited number of dynamic clusters and discrete sampling, which under-represents high-frequency fluctuations and thus yields a modestly elevated empirical LCR.

	\subsubsection{RMS Doppler Spread}

\begin{figure}[t]
		\centering
		\subfloat[]
		{
			\hspace{-1cm}
			\includegraphics[width=9.9cm]{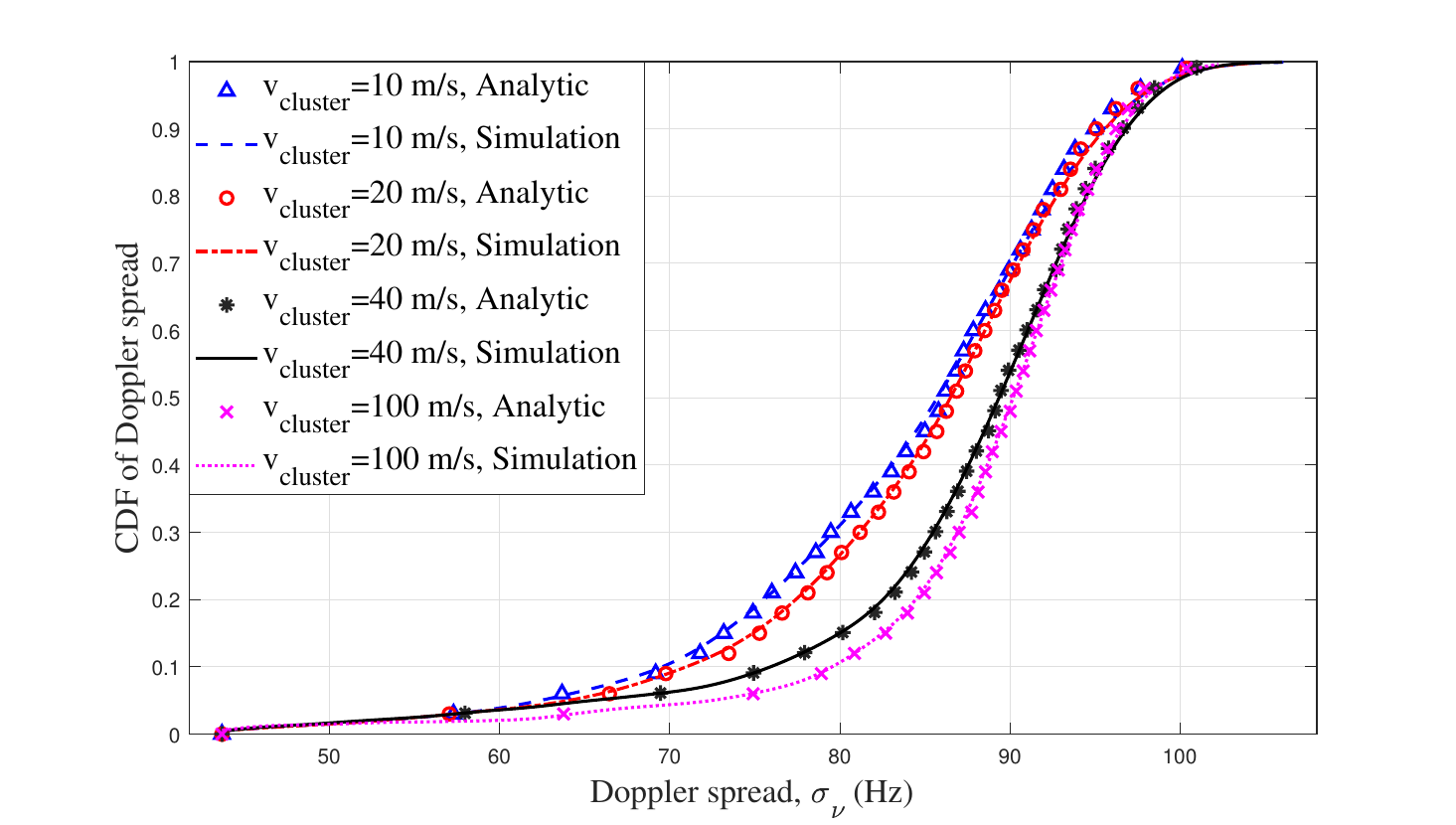}
			\label{fig:Doppler_clusterspeed}
		}\\
		\subfloat[]
		{
			\hspace{-1cm}
			\includegraphics[width=9.9cm]{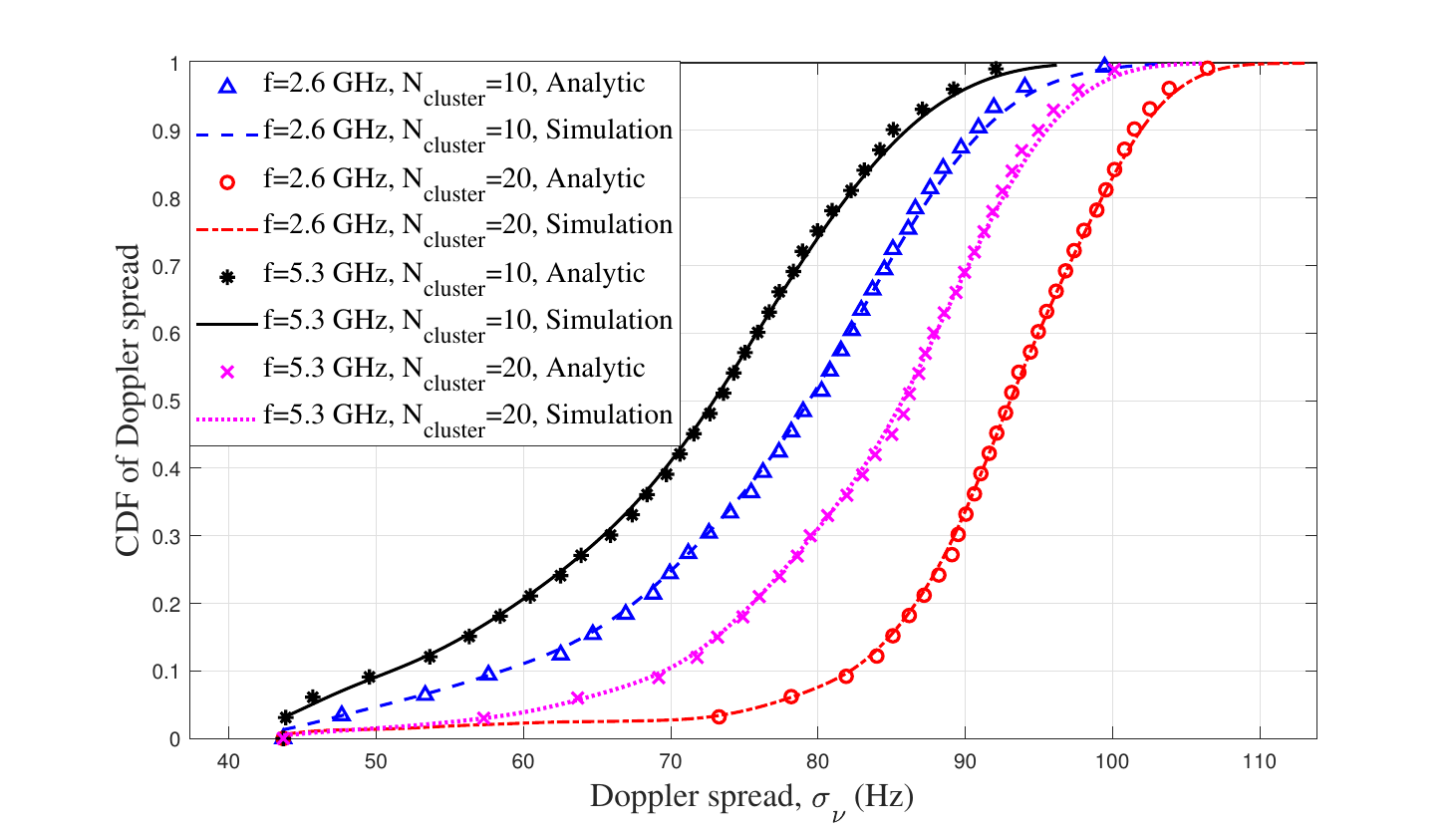}
			\label{fig:Doppler_f_clusternumber}
		}
		\caption{CDFs of RMS Doppler spread (a) with different cluster speeds and (b) at different frequencies with different dynamic cluster numbers.}
		\label{fig:CDF of Doppler spread}
		\vspace{-3mm}
	\end{figure}

		In Fig. \ref{fig:CDF of Doppler spread}\subref{fig:Doppler_clusterspeed}, we compare the CDFs of the RMS Doppler spread with different cluster speeds. It can be observed that the RMS Doppler spread increases with the dynamic cluster speed. The simulation result illustrates the larger channel Doppler shift caused by the higher mobility of clusters. Moreover, the analytical results match closely with the simulation curves, validating the accuracy of the analytical expressions.

		The comparison of the CDFs of RMS Doppler spread with different carrier frequencies and dynamic cluster numbers is shown in Fig. \ref{fig:CDF of Doppler spread}\subref{fig:Doppler_f_clusternumber}. The RMS Doppler spread increases with decreasing frequency. Because the degree of channel dispersion in the Doppler frequency domain increases when the carrier frequency becomes smaller. Additionally, Fig. \ref{fig:CDF of Doppler spread}\subref{fig:Doppler_f_clusternumber} demonstrates that the RMS Doppler spread increases significantly with the increasing number of dynamic clusters, mainly due to the more frequent motion of dynamic clusters. The analytical results again show good agreement with the simulation results, further confirming the effectiveness of our analytical modeling approach.

	\subsubsection{FCF}
	\begin{figure}[t]
		\centering
		\includegraphics[width=9cm]{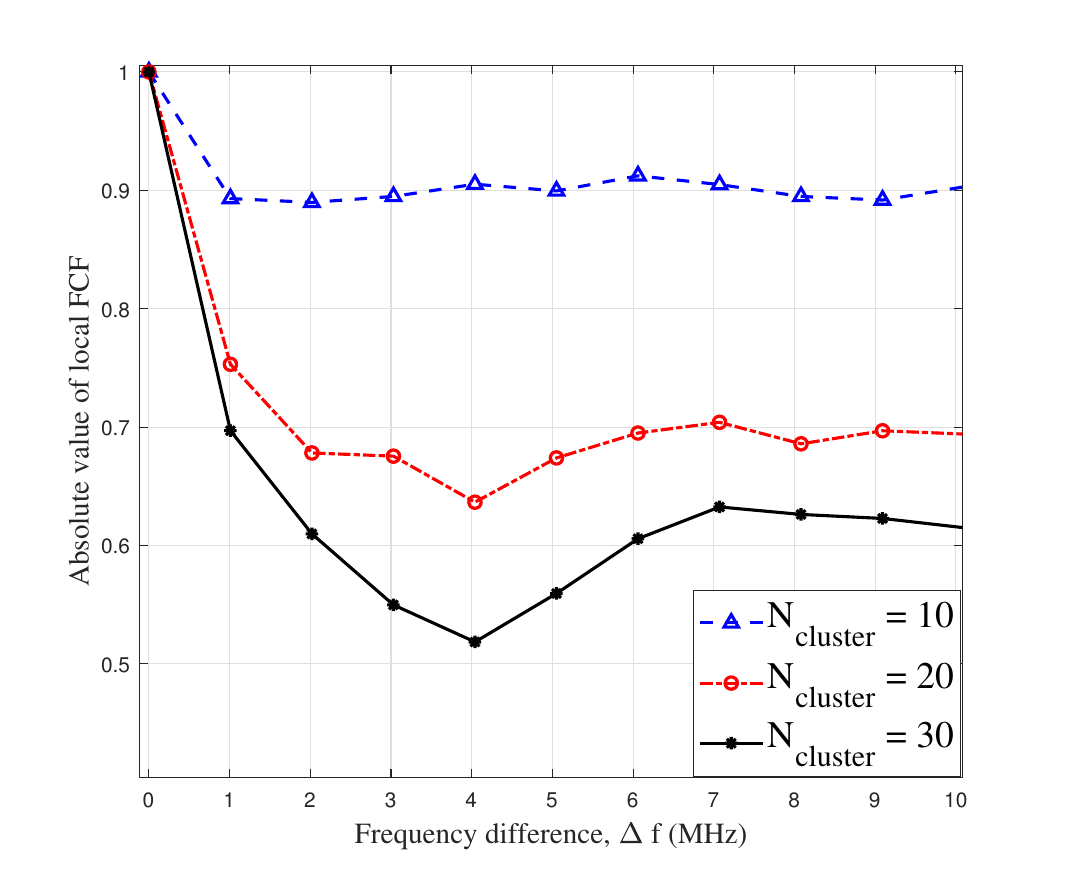}
		\caption{FCFs with different dynamic cluster numbers.}
		\label{fig:FCF_clusternumber}
		\vspace{-1mm}
	\end{figure}

	Fig. \ref{fig:FCF_clusternumber} illustrates that FCF becomes smaller with the increase of the dynamic cluster number. The FCF measures the frequency correlatio of the channel. It can be seen from \eqref{FCF} that FCF is mainly affected by the instantaneous total distance between the Tx, dynamic scatterers, static scatterers, and Rx. Consequently, by increasing the dynamic clusters, the travel distance of dynamic scattering channel increase significantly, leading to a decrease of FCF. The simulation results in Fig. \ref{fig:FCF_clusternumber} accord well with the analysis, validating the correctness of the RT-GSHCM.

\section{Conclusions}
In this paper, a novel DCM for 6G communications has been proposed. To construct the DCM both accurately and time-efficiently, we have developed a RT-GSHCM, modeling the static and dynamic IOs in physical environment by RT and 6GPCM, respectively. Furthermore, based on channel measurements and the reconstruction of static communication environment with RT, DCM in this area has been constructed. To validate the accuracy of DCM, the RMS angular spread, RMS delay spread, and delay PSD have been compared between the RT-GSHCM, channel measurements, RT, and 6GPCM. Meanwhile, the time cost on channel map updating has been compared between DCM and conventional channel maps. The results have shown that the channel information from DCM is accurate while the DCM comsumes much less time on update than conventional channel maps. Finally, through the RT-GSHCM, the statistical properties including RMS delay spread, RMS angular spread, delay PSD, LCR, RMS Doppler spread, and FCF have also been derived and analyzed with different model configurations. The simulation results have demonstrated that the channel variation brought by the change of model configurations is consistent well with the variation of physical environment. These results have verified the correctness and scalability of the RT-GSHCM.

\vspace{-12 mm}
\begin{IEEEbiography}
	[{\includegraphics[width=1in,height=1.25in,clip,keepaspectratio]{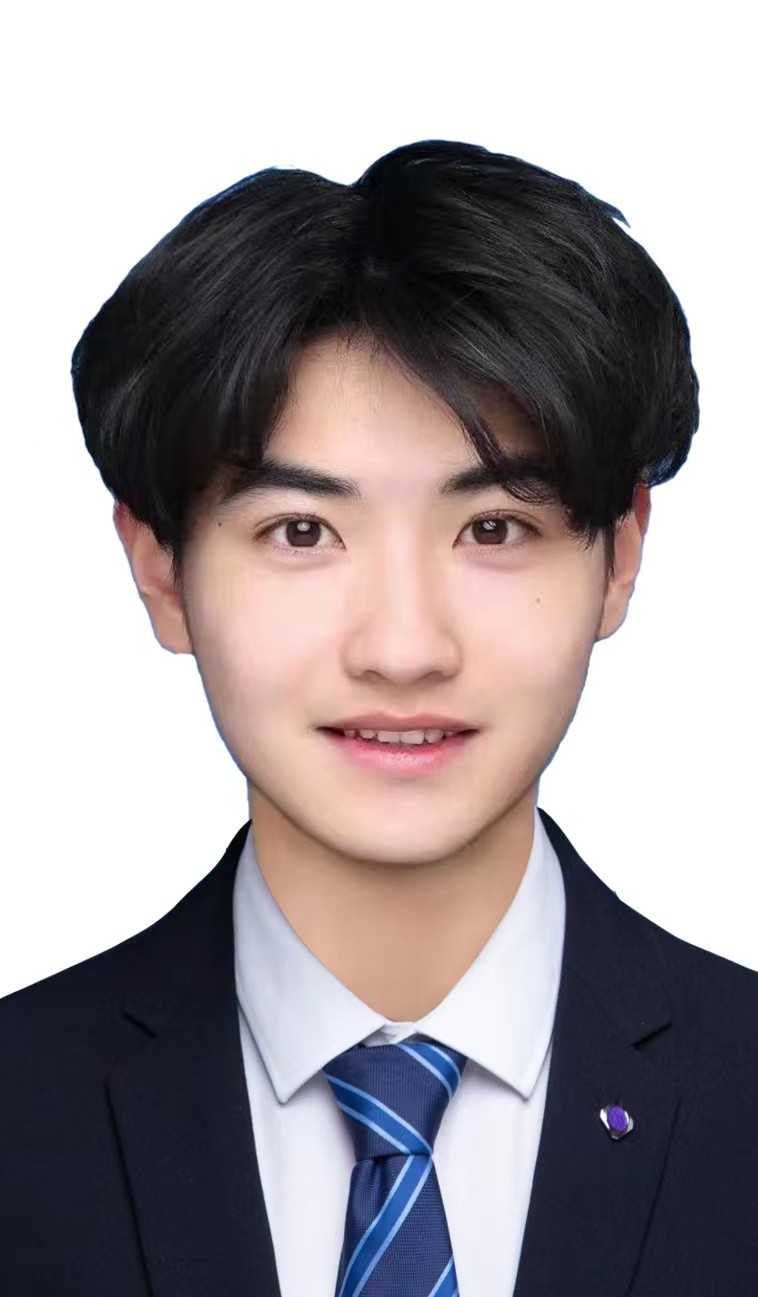}}]
	{Tianrun Qi} is currently pursuing a Ph.D. degree at the National Mobile Communications Research Laboratory, School of Information Science and Engineering, Southeast University, Nanjing 210096, China. He received a B.E. degree in information engineering from Southeast University, Nanjing, China, in 2022. His current research interests include 6G channel modeling and 6G dynamic channel maps. He is a Student Member of IEEE.
\end{IEEEbiography}

\vspace{-12 mm}
\begin{IEEEbiography}
	[{\includegraphics[width=1in,height=1.25in,clip,keepaspectratio]{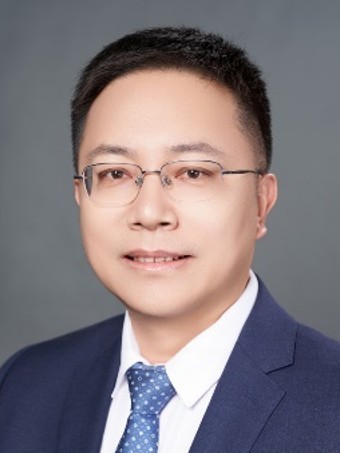}}]  					                
	{Cheng-Xiang Wang} (Fellow, IEEE) received the B.Sc. and M.Eng. degrees in communication and information systems from Shandong University, China, in 1997 and 2000, respectively, and the Ph.D. degree in wireless communications from Aalborg University, Denmark, in 2004.
	
	He was a Research Assistant with the Hamburg University of Technology, Hamburg, Germany, from 2000 to 2001, a Visiting Researcher with Siemens AG Mobile Phones, Munich, Germany, in 2004, and a Research Fellow with the University of Agder, Grimstad, Norway, from 2001 to 2005. He was with Heriot-Watt University, Edinburgh, U.K., from 2005 to 2018, where he was promoted to a professor in 2011. He has been with Southeast University, Nanjing, China, as a professor since 2018, and he is now the Dean of the School of Information Science and Engineering. He is also a professor with Pervasive Communication Research Center, Purple Mountain Laboratories, Nanjing, China. He has authored 4 books, 3 book chapters, and over 640 papers in refereed journals and conference proceedings, including 29 highly cited papers. He has also delivered 34 invited keynote speeches/talks and 23 tutorials in international conferences. His current research interests include wireless channel measurements and modeling, 6G wireless communication networks, and electromagnetic information theory.
	
	Dr. Wang is a Member of the Academia Europaea (The Academy of Europe), a Member of the European Academy of Sciences and Arts (EASA), a Fellow of the Royal Society of Edinburgh (FRSE), IEEE, and IET, an IEEE Communications Society Distinguished Lecturer in 2019 and 2020, a Highly-Cited Researcher recognized by Clarivate Analytics in 2017-2020. He is currently an Executive Editorial Committee Member of the \uppercase{IEEE Transactions on Wireless Communications}. He has served as an Editor for over sixteen international journals, including the \uppercase{IEEE Transactions on Wireless Communications}, from 2007 to 2009, the \uppercase{IEEE Transactions on Vehicular Technology}, from 2011 to 2017, and the \uppercase{IEEE Transactions on Communications}, from 2015 to 2017. He was a Guest Editor of the \uppercase{IEEE Journal on Selected Areas in Communications}, the \uppercase{IEEE Transactions on Big Data}, and the \uppercase{IEEE Transactions on Cognitive Communications and Networking}. He has served as a TPC Chair and General Chair for more than 30 international conferences. He received IEEE Neal Shepherd Memorial Best Propagation Paper Award in 2024. He also received 19 Best Paper Awards from international conferences.
\end{IEEEbiography}

\vspace{-12 mm}
\begin{IEEEbiography}
	[{\includegraphics[width=1in,height=1.25in,clip,keepaspectratio]{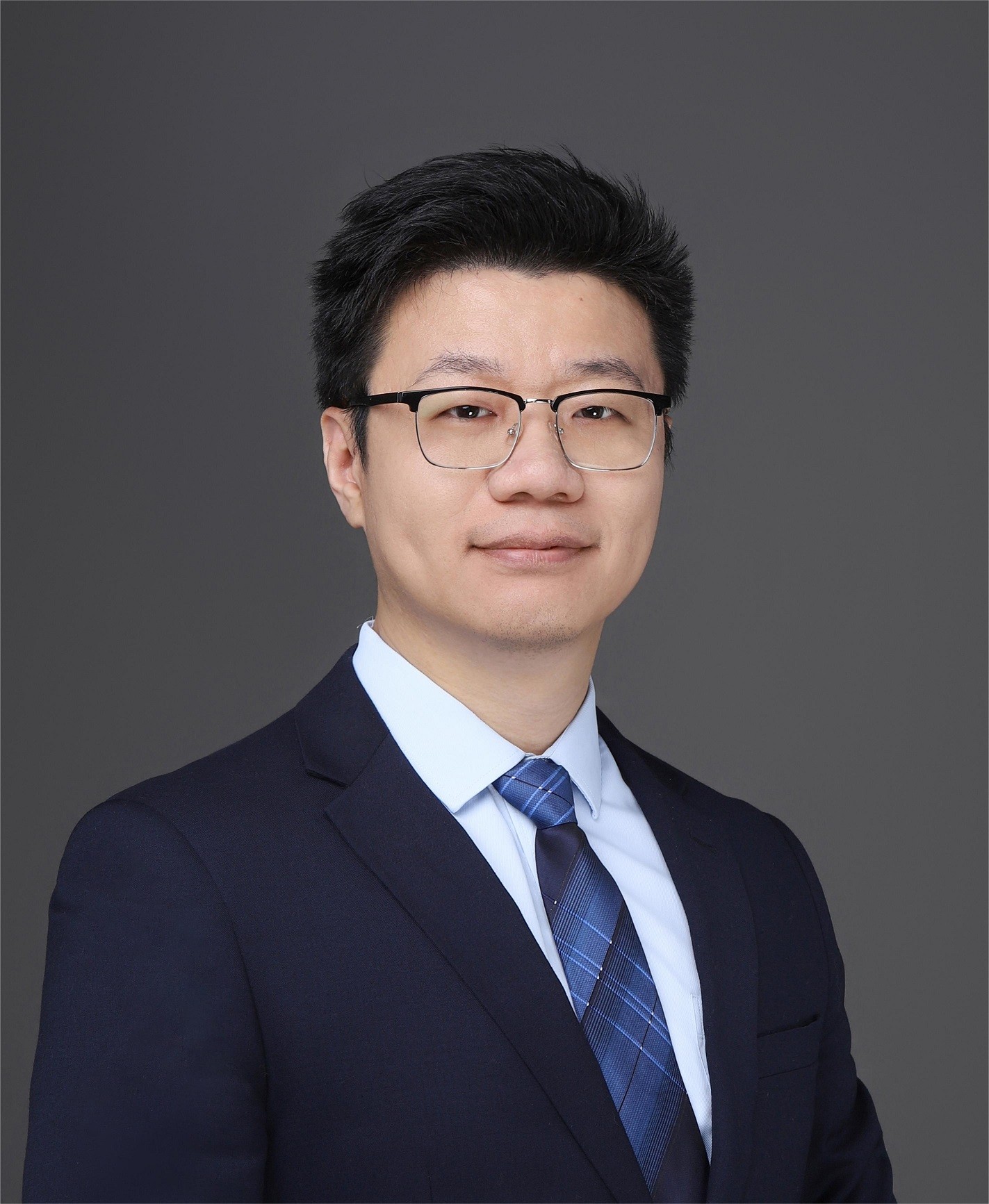}}] 					                                           
	{Chen Huang} (Member, IEEE) received the Ph.D. degree from Beijing Jiaotong University, Beijing, China, in 2021. From 2018 to 2020, he was a Visiting Scholar with the University of Southern California (USC), Los Angeles, CA, USA and was also with the Universite Catholique de Louvain (UCL), Louvain-la-Neuve, Belgium. From April 2021 to April 2023, he was a Postdoctoral Research Associate with Pervasive Communication Research Center, Purple Mountain Laboratories (PML), Nanjing, China, and was also with the National Mobile Communications Research Laboratory, School of Information Science and Engineering, Southeast University (SEU), Nanjing, China. Since April 2023, he has been an Research Associate Professor with Pervasive Communication Research Center, PML, and an Extramural Supervisor with the National Mobile Communications Research Laboratory, School of Information Science and Engineering, SEU, China. He has authored/co-authored one book chapters, more than 60 journal and conference papers, as well as 17 patents. His research interests include 6G channel measurements, characterization, and modeling, machine learning-based channel prediction, and localization. Dr. Huang was selected in Young Elite Scientists Sponsorship Program by China Association for Science and Technology and Outstanding Postdoctoral Fellow Program in Jiangsu, was the recipient of the four times the Best Paper Award from IEEE/CIC ICCC 2024, IEEE ICCT2023, WCSP 2018, and IEEE/CIC ICCC 2018. He is also the Associate Editor for IEEE Transactions on Vehicular Technology. He is the Technical Program Committee (TPC) Member for several conferences, including GlobeCom, ICC, VTC-fall, and VTC-spring.
\end{IEEEbiography}

\vspace{-12 mm}
\begin{IEEEbiography}
	[{\includegraphics[width=1in,height=1.25in,clip,keepaspectratio]{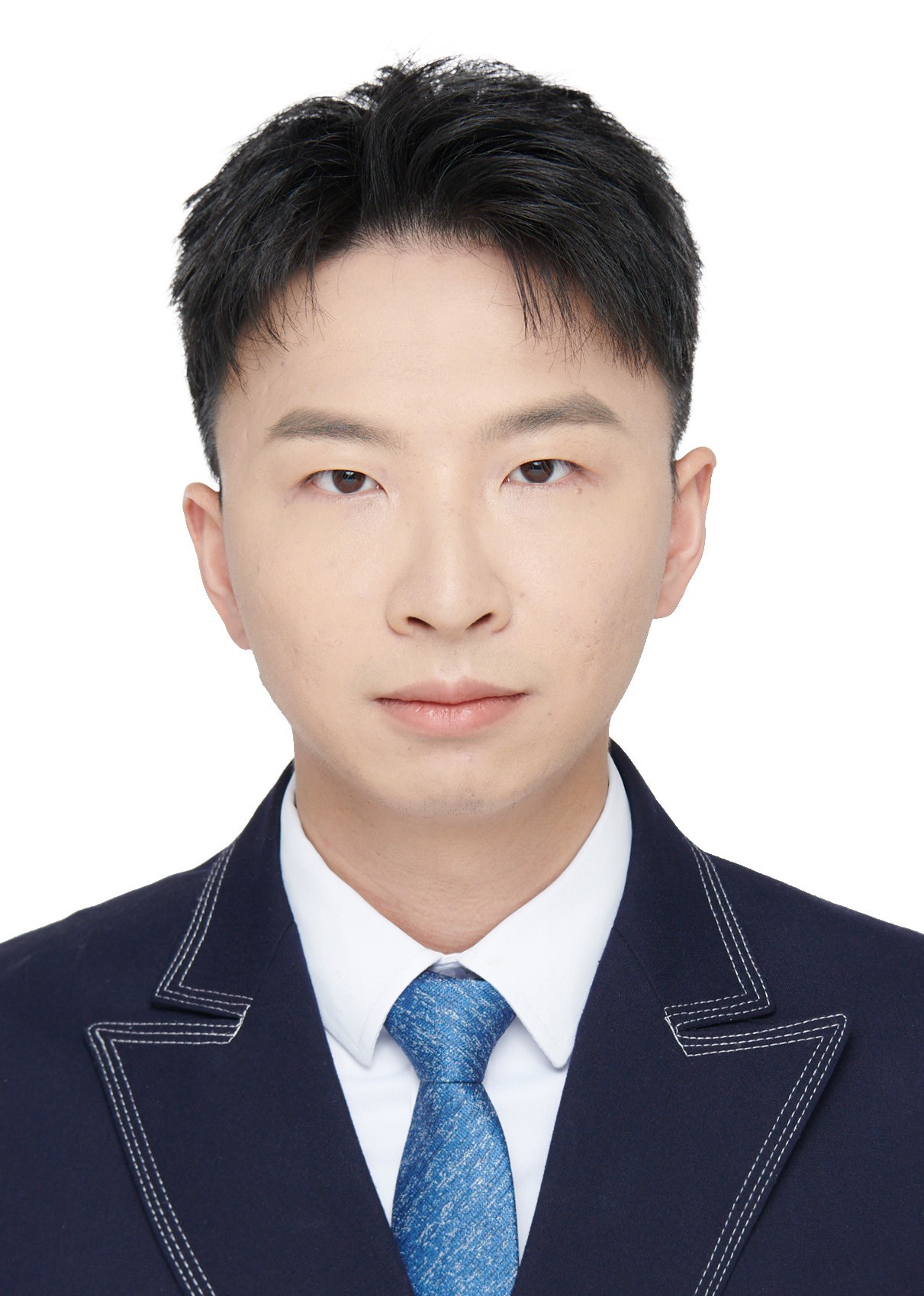}}] 		
	{Jiayue Shi} received the B.E. degree in Information Engineering from Southeast University, Nanjing, China, in 2022. He is currently pursuing the master's degree in the National Mobile Communications Research Laboratory, Southeast University, China. His research interests are 6G ray tracing channel modeling.
\end{IEEEbiography}

\vspace{-12 mm}
\begin{IEEEbiography}
	[{\includegraphics[width=1in,height=1.25in,clip,keepaspectratio]{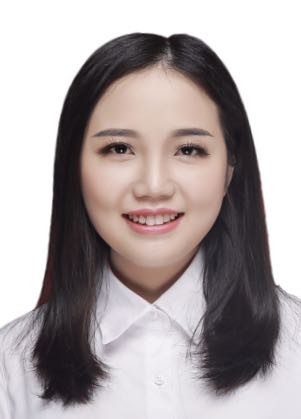}}] 		
	{Junling Li} (Member, IEEE) received the B.S. degree from Tianjin University, Tianjin, China, in 2013, the M.S. degree from Beijing University of Posts and Telecommunications, Beijing, China, in 2016, and the Ph.D. degree from the Department of Electrical and Computer Engineering, University of Waterloo, Waterloo, ON, Canada, in 2020.,She was a Joint Post-Doctoral Research Fellow at Shenzhen Institute of Artificial Intelligence and Robotics for Society (AIRS), Shenzhen, China; the University of Waterloo; and the Chinese University of Hong Kong, Shenzhen, from 2020 to 2022. She is currently an Associate Professor with the National Mobile Communications Research Laboratory, Southeast University, Nanjing, China. She has authored or co-authored more than 40 papers in refereed journals and conference proceedings. Her research interests include deep learning, image and video processing, digital twin online channel modeling, and 6G channel measurements and modeling.,Dr. Li received the Best Paper Awards from ICCC 2019 and ICCT 2023
\end{IEEEbiography}

\vspace{-12 mm}
\begin{IEEEbiography}
	[{\includegraphics[width=1in,height=1.25in,clip,keepaspectratio]{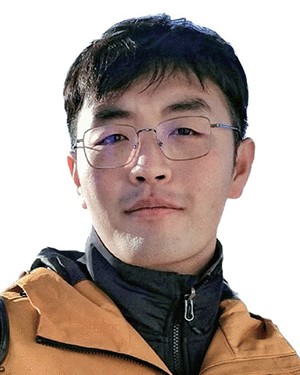}}] 		
	{Shuaifei Chen} (Member, IEEE) received the B.S. degree in communication engineering and the Ph.D. degree in information and communication engineering from Beijing Jiaotong University, Beijing, China, in 2018 and 2023, respectively. From 2019 to 2020, he visited the Department of Communication Systems, Linköping University, Linköping, Sweden. From 2021 to 2022, he visited the Division of Communication Systems, KTH Royal Institute of Technology, Stockholm, Sweden. He is currently a Research Fellow with the Pervasive Communication Research Center, Purple Mountain Laboratories, Nanjing, China, and also a Postdoctoral Researcher with the National Mobile Communications Research Laboratory, School of Information Science and Engineering, Southeast University, Nanjing. His research interests include signal processing and resource allocation for wireless communications, cell-free massive MIMO, and channel map-aided 6G multiple antenna technologies. He was recognized as an Exemplary Reviewer of IEEE Transactions on Communications in 2021 and IEEE Communications Letters in 2023.
\end{IEEEbiography}

\vspace{-12 mm}
\begin{IEEEbiography}
	[{\includegraphics[width=1in,height=1.25in,clip,keepaspectratio]{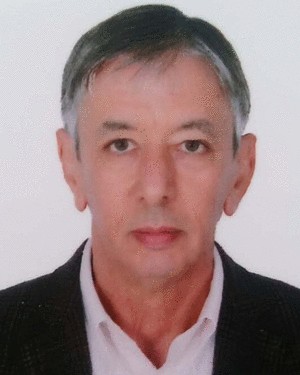}}] 		
	{El-Hadi M. Aggoune} (Life Senior Member, IEEE) received the M.S. and Ph.D. degrees in electrical engineering from the University of Washington (UW), Seattle, WA, USA, in 1984 and 1988, respectively. He is currently a Professor and the Director with the Sensor Networks and Cellular Systems Research Center, University of Tabuk, Tabuk, Saudi Arabia. He is listed as an inventor in several patents, one of them was assigned to Boeing Company, Chicago, IL, USA. He is a Professional Engineer registered in the state of Washington. He has coauthored articles in IEEE and other journals and conferences and was on editorial boards and technical committees for many of them. His research interests include wireless communication, sensor networks, power systems, neurocomputing, and scientific visualization. Dr. Aggoune was the recipient of the IEEE Professor of the Year Award from UW. He was the Director of the laboratory that received the Boeing Supplier Excellence Award. He was with several universities in the USA and abroad at many academic ranks, including endowed Chair Professor.
\end{IEEEbiography}	
\end{document}